\documentclass[11pt,a4paper]{article}
\pdfoutput=1

\usepackage{jheppub}
\usepackage{latexsym}
\usepackage{multirow}
\usepackage{color}
\usepackage[usenames,dvipsnames,table]{xcolor}
\usepackage{graphicx}
\usepackage{epsfig}  
\usepackage{epsf}    
\usepackage{dcolumn}
\usepackage{bm}
\usepackage{dcolumn}
\usepackage{textcomp}
\usepackage{float}
\usepackage{subfig}
\usepackage{hypcap}
\usepackage[]{hyperref}
\usepackage{makecell}
\usepackage{color}
\usepackage{pifont}
\usepackage{appendix}
\usepackage{amsmath}
\usepackage{multirow,bigdelim}  
\usepackage{lineno}
\usepackage[normalem]{ulem}

\hypersetup{
  bookmarks=true,         
  unicode=false,          
  pdftoolbar=true,        
 pdfmenubar=true,        
 pdffitwindow=true,     
 pdfstartview={FitH},    
 pdfsubject={Neutrino Oscillations Phenomenology},   
 pdfnewwindow=true,      
 pdfcreator={RevTeX},
 colorlinks=true,       
 linkcolor=red,          
 citecolor=blue,        
 filecolor=black,      
 urlcolor=blue,           
  }

\newcommand{\ldm}{\ensuremath{\Delta m_{31}^2}}
\newcommand{\sdm}{\ensuremath{\Delta m_{21}^2}}
\newcommand{\eem}{\ensuremath{\varepsilon_{e\mu}}}
\newcommand{\eet}{\ensuremath{\varepsilon_{e\tau}}}
\newcommand{\emt}{\ensuremath{\varepsilon_{\mu\tau}}}
\newcommand{\eee}{\ensuremath{\varepsilon_{ee}}}
\newcommand{\emm}{\ensuremath{\varepsilon_{\mu\mu}}}
\newcommand{\ett}{\ensuremath{\varepsilon_{\tau\tau}}}
\newcommand{\eff}{\ensuremath{\gamma-\beta}}
\newcommand{\txm}{\ensuremath{\theta_{12}^{m}}}
\newcommand{\tym}{\ensuremath{\theta_{13}^{m}}}
\newcommand{\tzm}{\ensuremath{\theta_{23}^{m}}}
\newcommand{\ldmm}{\ensuremath{\Delta m^2_{31,m}}}
\newcommand{\sdmm}{\ensuremath{\Delta m^2_{21,m}}}
\newcommand{\ie}{\textit{i.e.}}

\newcommand{\capdef}{}
\newcommand{\mycaption}[2][\capdef]{\renewcommand{\capdef}{#2}
	\caption[#1]{{\footnotesize #2}}}

\preprint{IP/BBSR/2021-1, CTPU-PTC-21-10}

\title{Evolution of Neutrino Mass-Mixing Parameters in Matter with Non-Standard Interactions}

\author[a,b,c]{Sanjib Kumar Agarwalla,}
\author[a,b]{Sudipta Das,}
\author[a,d]{Mehedi Masud,}
\author[a,b]{Pragyanprasu Swain} 

\affiliation[a]{Institute of Physics, Sachivalaya Marg, Sainik School Post, Bhubaneswar 751005, India}
\affiliation[b]{Homi Bhabha National Institute, Training School Complex, Anushakti Nagar, Mumbai 400094, India}
\affiliation[c]{International Centre for Theoretical Physics, Strada Costiera 11, 34151 Trieste, Italy}
\affiliation[d]{Center for Theoretical Physics of the Universe, Institute for Basic Science (IBS), Daejeon 34126, South Korea}	

\emailAdd{sanjib@iopb.res.in (ORCID: 0000-0002-9714-8866)} 
\emailAdd{sudipta.d@iopb.res.in (ORCID:0000-0002-5508-7751)} 
\emailAdd{masud@ibs.re.kr (ORCID: 0000-0002-7014-3520)} 
\emailAdd{pragyanprasu.s@iopb.res.in (ORCID: 0000-0003-3008-480X)}

\abstract
{We explore the role of matter effect in the evolution 
of neutrino oscillation parameters in the presence of 
lepton-flavor-conserving and lepton-flavor-violating 
neutral-current non-standard interactions (NSI) of 
the neutrino. We derive simple approximate 
analytical expressions showing the evolution/running 
of mass-mixing parameters in matter with energy in the 
presence of standard interactions (SI) and SI+NSI
(considering both positive and negative values of 
real NSI parameters). We observe that only the NSI 
parameters in the (2,3) block, namely $\varepsilon_{\mu\tau}$
and $(\gamma - \beta) \equiv (\varepsilon_{\tau\tau}
- \varepsilon_{\mu\mu})$ affect the running of 
$\theta_{23}$. Though all the NSI parameters influence 
the evolution of $\theta_{13}$, $\varepsilon_{e\mu}$ 
and $\varepsilon_{e\tau}$ show a stronger impact 
at the energies relevant for DUNE. The solar mixing angle 
$\theta_{12}$ quickly approaches to $\sim$ $90^{\circ}$ 
with increasing energy in both SI and SI+NSI cases.
The change in $\Delta m^2_{21,m}$ is quite significant
as compared to $\Delta m^2_{31,m}$ both in SI and 
SI+NSI frameworks for the energies relevant for DUNE 
baseline. Flipping the signs of the NSI parameters alters 
the way in which mass-mixing parameters run with energy. 
We demonstrate the utility of our approach in addressing 
several important features related to neutrino oscillation 
such as: a) unraveling interesting degeneracies between 
$\theta_{23}$ and NSI parameters, b) estimating the 
resonance energy in presence of NSI when $\theta_{13}$ 
in matter becomes maximal, c) figuring out the required 
baselines and energies to have maximal matter effect 
in $\nu_{\mu}$ $\rightarrow$ $\nu_{e}$ transition in the 
presence of different NSI parameters, and d) studying 
the impact of NSI parameters $\varepsilon_{\mu\tau}$ 
and $(\gamma - \beta)$ on the $\nu_{\mu} \to \nu_{\mu}$ 
survival probability.}

\arxivnumber{2103.aaaaa}

\begin{document}
\maketitle
\flushbottom

\section{Introduction and Motivation}

The phenomenon of three-flavor neutrino oscillation
is governed by the six fundamental mass-mixing 
parameters~\cite{Zyla:2020zbs}:
a) three mixing angles: $\theta_{12}, \theta_{13}, \theta_{23}$,
b) two independent mass-squared differences: 
$\sdm \equiv m^{2}_{2} - m^{2}_{1}$, $\ldm \equiv m^{2}_{3} - m^{2}_{1}$,
and c) one Dirac CP phase $\delta_{\mathrm{CP}}$.  
After the discovery of neutrino oscillation at the 
Super-Kamiokande (Super-K) experiment in 1998~\cite{Fukuda:1998mi},
fantastic data from the world-class accelerator, atmospheric, 
reactor, and solar neutrino experiments are pouring in day-by-day
to commence the era of precision neutrino measurement 
science~\cite{Marrone:2021,NuFIT,Esteban:2020cvm,deSalas:2020pgw}, 
which will certainly provide crucial insights on the possible origin of 
neutrino mass and mixing~\cite{Mohapatra:2005wg,Strumia:2006db,GonzalezGarcia:2007ib}.

Marvelous data from several ongoing experiments such as
Super-K~\cite{Ashie:2004mr}, IceCube-DeepCore~\cite{Aartsen:2017nmd}, 
ANTARES~\cite{Albert:2018mnz}, Daya Bay~\cite{Adey:2018zwh}, RENO~\cite{Ahn:2012nd},  
Tokai to Kamioka (T2K)~\cite{Abe:2019vii,Abe:2021gky}, and 
NuMI Off-axis $\nu_e$ Appearance (NO${\nu}$A)~\cite{Acero:2019ksn} 
have been improving our knowledge about the neutrino oscillation 
parameters beyond expectations. Because of this fascinating 
progress, we have been able to build a robust, simple, three-flavor 
neutrino oscillation paradigm which successfully accommodate 
most of the data~\cite{Marrone:2021,NuFIT,Esteban:2020cvm,deSalas:2020pgw}.

Future high-precision neutrino oscillation experiments such as 
the Deep Underground Neutrino Experiment (DUNE)~\cite{Abi:2020evt,Abi:2021arg}, 
Tokai to Hyper-Kamiokande (T2HK)~\cite{Abe:2015zbg}, Tokai to Hyper-Kamiokande 
with a second detector in Korea (T2HKK)~\cite{Abe:2016ero}, 
European Spallation Source $\nu$ Super Beam (ESS$\nu$SB)~\cite{Baussan:2013zcy}, 
India-based Neutrino Observatory (INO)~\cite{Devi:2014yaa,Kumar:2017sdq,Kumar:2020wgz}, 
Jiangmen Underground Neutrino Observatory (JUNO)~\cite{An:2015jdp}, 
and THEIA~\cite{Askins:2019oqj} aim to determine the oscillation parameters 
with a precision around a {\it few} \%. Therefore, these next generation experiments
are potentially sensitive to various sub-leading beyond the Standard Model (BSM)
effects~\cite{Arguelles:2019xgp,Agarwalla:2020}. One such interesting BSM 
scenario is non-standard neutrino interactions 
(NSI)~\cite{Wolfenstein:1977ue,Valle:1987gv,Guzzo:1991hi,Roulet:1991sm,Grossman:1995wx,Guzzo:2000kx,Huber:2001zw,Gago:2001si,Escrihuela:2011cf,GonzalezGarcia:2011my,Ohlsson:2012kf,Gonzalez-Garcia:2013usa,Miranda:2015dra,Farzan:2017xzy,Khatun:2019tad,Dev:2019anc,Kumar:2021lrn} 
which is the main focus of this paper. 

Analytical understanding of neutrino oscillation probabilities 
over a wide range of energies and baselines becomes non-trivial 
in the presence of standard interactions (SI)\footnote{They 
appear into the picture due to the Standard Model (SM) 
$W$-exchange interactions between the ambient matter 
electrons and the propagating electron neutrinos, 
which is popularly known as the 
`MSW effect'~\cite{Wolfenstein:1977ue,Mikheev:1986gs,Mikheev:1986wj}.}.
Now, on top of that if NSI exit in Nature then the task becomes 
even more complex. Assuming the line-averaged constant Earth matter
density for a given baseline, several authors have derived approximate
analytical expressions for the neutrino oscillation 
probabilities\footnote{In Ref.~\cite{Parke:2019vbs}, 
the authors performed a detailed comparative study between
different expansions for neutrino oscillation probabilities 
in the presence of SI in matter. They also studied the accuracy 
and computational efficiency of several exact and approximate 
expressions for neutrino oscillation probabilities in the context
of long-baseline (LBL) experiments.} 
in the presence of SI~\cite{Petcov:1986qg,Kim:1986vg,Arafune:1996bt,Arafune:1997hd,Ohlsson:1999xb,Freund:2001pn,Cervera:2000kp,Akhmedov:2004ny,Asano:2011nj,Agarwalla:2013tza,Minakata:2015gra,Denton:2016wmg} and SI+NSI~\cite{GonzalezGarcia:2001mp,Ota:2001pw,Yasuda:2007jp,Kopp:2007ne,Ribeiro:2007ud,Blennow:2008eb,Kikuchi:2008vq,Meloni:2009ia,Agarwalla:2015cta}.

To obtain a better understanding of the neutrino oscillation 
probabilities as functions of baseline $L$ and/or neutrino 
energy $E$ in the presence of SI or SI+NSI, it is quite
important in the first place to have a clear knowledge 
on how various mixing angles and mass-squared 
differences get modified in matter with energy for
a given baseline. Simple approximate analytical
expressions showing the evolution/running of 
mass-mixing parameters in matter with energy
in the presence of SI and SI+NSI allow us to
address several important features that show
up in neutrino oscillation in a more general and
transparent fashion. This simple and more intuitive way 
to understand the neutrino oscillation phenomena 
will likely pave a way to disentangle the various 
non-trivial correlations/degeneracies that may
be present among the various oscillation and 
NSI parameters. This paper addresses several
pressing issues along this direction.

There exist several studies in the literature
investigating how the presence of SI and NSI
affect the evolution of effective neutrino 
oscillation parameters (the mixing angles, 
mass-squared differences, and CP-violating 
phase) in matter with energy, and eventually
how they modify the oscillation probabilities~\cite{Barger:1980tf,Zaglauer:1988gz,Ohlsson:1999um,Freund:2001pn,Kimura:2002hb,Kimura:2002wd,Akhmedov:2004ny,Kikuchi:2008vq,Meloni:2009ia,Agarwalla:2013tza,Minakata:2015gra,Agarwalla:2015cta,Denton:2016wmg}.
In Refs.~\cite{Barger:1980tf,Zaglauer:1988gz,Kimura:2002hb,Kimura:2002wd},
the authors diagonalize analytically the three-flavor 
propagation Hamiltonian in constant-density matter 
to obtain the exact expressions for the modified 
mass-mixing parameters in the presence of SI. 
The authors in Ref.~\cite{Ohlsson:1999um} make
use of the Cayley-Hamilton approach with a plane
wave approximation to derive the expressions for
the modified mass-mixing parameters without 
performing the actual diagonalization of the 
Hamiltonian. They also briefly discuss how
these oscillation parameters get modified 
with the strength of SI. In Ref.~\cite{Freund:2001pn}, 
the author diagonalizes the neutrino propagation 
Hamiltonian in the presence of SI by applying 
successive rotations and obtain the expressions
for the modified mass-mixing parameters.
In Ref.~\cite{Kimura:2002wd}, the authors make
use of the relations between the Jarlskog invariants 
in vacuum and matter (Naumov-Harrison-Scott 
identities~\cite{Jarlskog:1985ht,Naumov:1991ju,Harrison:1999df})
to derive the expressions for modified mass-mixing parameters 
in the presence of SI in constant-density matter.
In Ref.~\cite{Meloni:2009ia}, the authors adopt
a perturbative approach towards the SI and NSI
effects and discuss the possible modifications
of the mass-mixing parameters. In most of these
studies, the authors extract the expressions for
modified mass-mixing parameters in order to 
obtain approximate analytical expressions for 
the neutrino oscillation probabilities. Using the
Jacobi method~\cite{Jacobi:1846}, the authors 
in Ref.~\cite{Agarwalla:2013tza} show that the
matter effect on neutrino oscillation due to SI 
could be assimilated into the evolution of the
effective mixing angles $\theta_{12}$ and 
$\theta_{13}$, and the effective mass-squared
differences in matter as functions of the 
Wolfenstein matter term $2\sqrt{2}G_F N_e E$, 
while the effective values of $\theta_{23}$ and 
$\delta_{\mathrm{CP}}$ remain unaltered. 
Here, $G_F$ is the Fermi muon decay constant,
$N_e$ is the ambient electron number density, 
and $E$ is the energy of the neutrino. They obtain 
the approximate neutrino oscillation probabilities
by simply replacing the mass-mixing parameters 
in the expressions for the probabilities in vacuum
with their running in-matter counterparts. Similar
approach is adopted by the authors in 
Ref.~\cite{Agarwalla:2015cta} to show the
evolution of mass-mixing parameters in the 
presence of lepton-flavor-conserving, 
non-universal NSI of the neutrino.

In the present work, we perform successive
rotations to almost diagonalize the propagation 
Hamiltonian in the presence of SI and SI+NSI 
and derive simple approximate analytical 
expression for the effective mass-mixing 
parameters in constant-density matter.
While deriving our expressions, we retain 
the terms of all orders in $\sin\theta_{13}$
and $\alpha$ (the ratio of solar and 
atmospheric mass-squared differences, 
$\Delta m^2_{21}/\Delta m^2_{31}$) which
are quite important in light of the large 
value of $\theta_{13}$. In our study, 
we also entertain all possible allowed 
values of $\theta_{23}$ in vacuum. 
As far as NSI are concerned, we consider 
all possible lepton-flavor-conserving 
and lepton-flavor-violating neutral-current (NC)
NSI at-a-time in our analysis which affect 
the propagation of neutrino in matter. 
We discuss many salient features of
the evolution of oscillation parameters
with energy for some benchmark choices
of baseline and study in detail how these
mass-mixing parameters get affected by
various combinations of NSI parameters.
Our simple analytical expressions enables 
us to explore the possible degeneracies 
between $\theta_{23}$ (which still has
large uncertainty) and NSI parameters
for a given choice of neutrino mass ordering 
in a simple manner. For the first time,
we show how the famous MSW-resonance
condition ($\theta_{13}$ in matter becomes 
$45^{\circ}$)~\cite{Wolfenstein:1977ue,Wolfenstein:1979ni,Mikheev:1986gs,Mikheev:1986wj}
gets altered in the presence of NC-NSI.
We demonstrate how the simple approximate
analytical expressions for the running of oscillation
parameters in matter help us to estimate 
the baselines and energies for which we 
have the maximal matter effect in 
$\nu_{\mu} \to \nu_{e}$ oscillation channel
in the presence of various NSI parameters.
For simplicity, we perform our calculations
in a CP-conserving scenario where the 
standard Dirac CP phase $\delta_{\mathrm{CP}}$ 
and the phases associated with the 
lepton-flavor-violating NSI parameters 
are assumed to be zero. We consider
both positive and negative values of 
real NSI parameters in our analysis.

We plan this paper in the following fashion. 
We start Sec.~\ref{sec:basics} with a brief
discussion on the theoretical formalism of
NSI. This is followed by a short summary 
of the existing bounds on the NC-NSI. 
In Sec.~\ref{sec:nsi}, we describe our 
method of approximately diagonalizing 
the effective neutrino propagation Hamiltonian 
in the presence of all possible NC-NSI in 
constant-density matter. Subsequently, 
we derive the expressions for the modified 
mass-mixing parameters. In Sec.~\ref{sec:angle_running}, 
we study the evolution of $\theta_{23}$,
$\theta_{13}$, and $\theta_{12}$ in matter
with energy in detail for some benchmark 
choices of baseline and analyze the role
of various NSI parameters on their running.
We illustrate the impact of SI and various 
NSI parameters on the running of two 
modified mass-squared differences 
in Sec.~\ref{sec:mass_running}. 
In Sec.~\ref{sec:resonance}, 
using the expressions for modified 
mass-mixing parameters, we 
estimate for the first time a simple
and compact expression for the
$\theta_{13}$-resonance energy
in the presence of all possible
NC-NSI parameters and identify
the NSI parameters that significantly
affect the $\theta_{13}$-resonance 
energy. We devote Sec.~\ref{sec:omsd}
to exhibit the utility of our approach 
in determining the baselines and energies
for which we can achieve the maximal
matter effect in $\nu_{\mu} \to \nu_{e}$
transition in the presence of various
NSI parameters. Section~\ref{sec:muon-survival-probability}
describes how the NSI parameters 
in the (2,3) block affect 
$\nu_{\mu} \to \nu_{\mu}$ 
disappearance channel.
Finally, we summarize and 
draw our conclusions in 
Sec.~\ref{sec:summary}.

\section{Theoretical Formalism of NSI}
\label{sec:basics}

NSI which arise naturally in most of the neutrino mass models 
can be of charged-current (CC) or neutral-current (NC) in nature. 
Both of them can be described with a dimension-six operator in 
the four-fermion effective Lagrangian~\cite{Wolfenstein:1977ue, Grossman:1995wx, Ohlsson:2012kf},
\begin{align}
&\mathcal{L}_{\mathrm{NC-NSI}} = -2\sqrt{2} G_{F} \sum_{\alpha, \beta, f, C} \varepsilon^{fC}_{\alpha\beta} ({\Bar\nu_{\alpha}}\gamma^{\mu}P_{L}\nu_{\beta}) (\Bar{f}\gamma_{\mu}P_Cf), 
\label{eq:Lnsi} \\
&\mathcal{L}_{\mathrm{CC-NSI}} = -2\sqrt{2} G_{F} \sum_{\alpha, \beta, f^{\prime}, f, C} \varepsilon^{ff^{\prime}C}_{\alpha\beta} ({\Bar\nu_{\alpha}}\gamma^{\mu}P_{L}l_{\beta}) (\Bar{f}^{\prime}\gamma_{\mu}P_Cf), 
\label{eq:Lnsi_CC}
\end{align}
where, $P_{C}$ indicates the chiral projection
operators $P_{L}$ or $P_{R}$. The dimensionless  
coefficients $\varepsilon^{fC}_{\alpha\beta}$ in
Eq.~\ref{eq:Lnsi} denote the strength of NC-NSI 
between the leptons of flavors $\alpha$ and $\beta$
($\alpha, \beta = e, \mu, \tau$), and the first 
generation fermions $f \in \{e, u, d\}$.
In Eq.~\ref{eq:Lnsi_CC}, the dimensionless
coefficients $\varepsilon^{ff^{\prime}C}_{\alpha\beta}$
indicate the strength of CC-NSI between the leptons 
of $\alpha$ and $\beta$ flavors ($\alpha, \beta = e, \mu, \tau$),
and the first generation fermions $f \neq f^{\prime} \in \{u, d\}$.
The hermiticity of these interactions imposes the following conditions:
\begin{equation}
\varepsilon_{\alpha\beta}^{fC} \;=\; (\varepsilon_{\beta\alpha}^{fC})^* \;, \;\;\;\;\;\;
\varepsilon_{\alpha\beta}^{ff^{\prime}C} \;=\; (\varepsilon_{\beta\alpha}^{ff^{\prime}C})^* \;.
\end{equation}

The CC-NSI modify the production and detection of neutrinos 
and may also lead to charged-lepton flavor violation. The NC-NSI, 
on the other hand, affect the propagation of neutrinos. Since the 
coupling strength $\varepsilon^{fC}_{\alpha\beta}$ enters into 
the Lagrangian only through vector coupling, we can write 
$\varepsilon^f_{\alpha\beta} = \varepsilon_{\alpha\beta}^{fL} + \varepsilon_{\alpha\beta}^{fR}$. 
It is worthwhile to mention here that models employing scalar mediators~\cite{Ge:2018uhz} 
or other spin structures~\cite{AristizabalSierra:2018eqm} are also available in the literature. 
Beyond a simplified model approach, many UV complete models for NSI have also been explored 
(see, for instance, \cite{Heeck:2011wj, Farzan:2015hkd, Farzan:2016wym, Babu:2017olk, Wise:2018rnb}).
For a recent comprehensive review of the NSI, see~\cite{Farzan:2017xzy}.  
Using Eqs.~\ref{eq:Lnsi} and \ref{eq:Lnsi_CC} and the well-known relation 
$G_{F}/\sqrt{2} \simeq g_{W}^{2}/8m_{W}^{2}$, it can be shown that the 
effective NSI parameters ($\varepsilon$) are proportional to  
$m_{W}^{2}/m_{X}^{2}$~\cite{GonzalezGarcia:2001mp,Kopp:2007ne,Minakata:2008gv},
where $g_{W}$ is the coupling constant of the weak interaction, $m_{W}$ is the W boson mass ($\simeq 80$ GeV $\sim 0.1$ TeV), 
and $m_{X}$ is the mass scale where NSI are generated. Thus it can easily be observed that for $m_{X} \sim 1$ TeV, the NSI 
parameters are of the order of $10^{-2}$.
	
In the present work, we concentrate on the NC-NSI 
which appear during neutrino propagation through
matter. Here, the effective NSI parameter can be 
written in the following fashion
\begin{equation}
\varepsilon_{\alpha\beta}
\;\equiv\;
\sum_{f=e,u,d}
\varepsilon_{\alpha\beta}^{f}
\dfrac{N_f}{N_e}
\;\equiv\;
\sum_{f=e,u,d}
\left(
\varepsilon_{\alpha\beta}^{fL}+
\varepsilon_{\alpha\beta}^{fR}
\right)\dfrac{N_f}{N_e}
\;.
\label{eq:eps_ab}
\end{equation}
Here, $N_{f}$ is the first generation ($e, u, d$) 
fermion number density in the ambient medium. 

The effective Hamiltonian for neutrinos propagating 
in matter in presence of all the lepton-flavor-conserving 
and lepton-flavor-violating NC-NSI can be written as
\begin{equation}
\label{eq:Heff}
H_f = \frac{1}{2E}\left[U\begin{pmatrix}
0&0&0\\
0&\Delta m^2_{21}&0\\
0&0&\Delta m^2_{31}\\
\end{pmatrix}U^{\dagger}+2EV_{CC}\begin{pmatrix}
1+\varepsilon_{ee}&\varepsilon_{e\mu}&\varepsilon_{e\tau}\\
\varepsilon^{\ast}_{e\mu}&\varepsilon_{\mu\mu}&\varepsilon_{\mu\tau}\\
\varepsilon^{\ast}_{e\tau}&\varepsilon^{\ast}_{\mu\tau}&\varepsilon_{\tau\tau}\\
\end{pmatrix}\right],
\end{equation}
where,
$\Delta m^2_{21}(\equiv m^2_2 - m^2_1)$ and 
$\Delta m^2_{31}(\equiv m^2_3 - m^2_1)$ are 
the solar and atmospheric mass-squared differences, 
respectively. $U$ is the $3 \times 3$ unitary Pontecorvo-Maki-Nakagawa-Sakata 
(PMNS) matrix in vacuum~\cite{Pontecorvo:1957qd,Maki:1962mu,Pontecorvo:1967fh},
which can be parametrized using the three mixing angles: $\theta_{12}$, 
$\theta_{23}$, $\theta_{13}$, and one Dirac-type CP phase $\delta_{\mathrm{CP}}$ 
(ignoring Majorana phases) in the following fashion
\begin{equation}
U =  R_{23}(\theta_{23},0)\;R_{13}(\theta_{13},\delta_{\mathrm{CP}})\;R_{12}(\theta_{12},0) \;.
\label{eq:U-parametrization}
\end{equation}
In Eqn~\ref{eq:Heff}, $V_{CC}$ is the standard 
$W$-exchange interaction potential in matter 
which can be expressed as
\begin{align}
\label{eq:Vcc}
V_{CC} \;=\; \sqrt{2}G_FN_{e}\approx 7.6 \times Y_{e} \times 10^{-14} 
\left[\frac{\rho_{\text{avg}}}{\mathrm{g/cm^3}}\right] \; \mathrm{eV} \;,
\end{align}
where $Y_{e} \;=\; N_e/(N_p+N_n)$ 
is the relative electron number density 
of the medium and $\rho_{\text{avg}}$ 
is the line-averaged constant matter density.
For the Earth matter which is the focus of our paper, 
it is safe to assume neutral and isoscalar matter, 
\textit{i.e.} $N_n\approx N_p = N_e$. Under these 
assumptions, the relative electron number 
density inside the Earth turns out to be 
$Y_{e} \approx 0.5$. 

The (1,1) element of the effective Hamiltonian 
$H_f$ (see Eq.~\ref{eq:Heff}) contains the term
$\varepsilon_{ee}V_{CC}$ which gets simply 
added to the standard matter effect term. 
Since it can mimic the role of standard 
interaction, it is a wise choice to subtract 
a common physical phase $I (\equiv \eee V_{CC})$ 
from the right-hand side (R.H.S.) of Eq.\ \ref{eq:Heff}. 
Then, the effective Hamiltonian takes the form
\begin{equation}
\label{eq:Heff1}
H_{f}=\Delta_{31}\left[U\begin{pmatrix}
0&0&0\\
0&\alpha&0\\
0&0&1\\
\end{pmatrix}U^{\dagger}+\hat{A}\begin{pmatrix}
1&\varepsilon_{e\mu}&\varepsilon_{e\tau}\\
\varepsilon^{\ast}_{e\mu}&\beta&\varepsilon_{\mu\tau}\\
\varepsilon^{\ast}_{e\tau}&\varepsilon^{\ast}_{\mu\tau}&\gamma\\
\end{pmatrix}\right] \;,
\end{equation}
where, $\Delta_{31} \equiv \Delta m^2_{31}/2E$,
$\alpha \equiv \Delta m^2_{21}/\Delta m^2_{31}$,
$\hat{A} \equiv 2EV_{CC}/\Delta m^2_{31}$.
We define the effective lepton-flavor-conserving
diagonal NC-NSI parameters as 
$\beta \equiv \emm - \eee$ 
and $\gamma \equiv \ett - \eee$.

\begin{table}[htb!]
\centering
\begin{tabular}{|c|c|}
\hline\hline
NSI parameters & $2\sigma$ bounds \\ \hline\hline
$\varepsilon_{e\mu}$ & $[-0.372, +0.301]$ \\ \hline
$\varepsilon_{e\tau}$ & $[-1.657, +0.732]$ \\ \hline 
$\varepsilon_{\mu\tau}$ & $[-0.076,+0.058]$ \\ \hline
$\beta$ $(\varepsilon_{\mu\mu}-\varepsilon_{ee})$ & $[-2.861,+0.144]$ \\ \hline 
$\gamma$ $(\varepsilon_{\tau\tau}-\varepsilon_{ee})$ & $[-2.892,+0.836]$ 
\\ \hline\hline
\end{tabular}
\mycaption{Bounds on the effective NC-NSI parameters 
from the neutrino oscillation experiments at 2$\sigma$ 
confidence level. Values of $\varepsilon^{f}_{\alpha\beta}$ 
in Eq.\ \ref{eq:eps} is taken from the global fit 
analysis~\cite{Esteban:2018ppq}.}
\label{tab:bound}	
\end{table}

We now briefly discuss the present constraints 
on the effective NC-NSI parameters obtained
from the global fit of neutrino oscillation 
data~\cite{Esteban:2018ppq}. 
Using Eq.\ \ref{eq:eps_ab}, we can write, 
\begin{align}
\varepsilon_{\alpha\beta}&= \varepsilon^{p}_{\alpha\beta} + Y_n \varepsilon^{n}_{\alpha\beta} \nonumber \\
& = (2 + Y_n)\varepsilon^u_{\alpha\beta}+(1+2Y_n)\varepsilon^d_{\alpha\beta} \;,
\label{eq:eps}
\end{align}
where, $Y_n$ is the average neutron/proton ration inside the Earth.
According to Ref.~\cite{Esteban:2018ppq}, $Y_n = 1.051$. 
Here, we have taken into account the fact that 
$N_{u} = 2N_{p} + N_{n}$ and $N_{d} = N_{p} + 2N_{n}$, 
which in turn imply that 
$\varepsilon_{\alpha\beta}^{p} = 2\varepsilon_{\alpha\beta}^{u} + \varepsilon_{\alpha\beta}^{d}$ 
and $\varepsilon_{\alpha\beta}^{n} = \varepsilon_{\alpha\beta}^{u} + 2\varepsilon_{\alpha\beta}^{d}$. 
Note that the contribution from $\varepsilon^e_{\alpha\beta}$ is not considered 
in the global $3\nu$ analysis in the presence of NC-NSI parameters~\cite{Esteban:2018ppq}. 
Now, we use the bounds ($2\sigma$) on $\varepsilon_{\alpha\beta}^{u}$ and $\varepsilon_{\alpha\beta}^{d}$ 
from the global fit analysis~\cite{Esteban:2018ppq} and list the subsequent $2\sigma$ bounds 
on the effective NSI parameters $\varepsilon_{\alpha\beta}$ in Table~\ref{tab:bound}.

\section{Diagonalization of the Effective Hamiltonian in the presence of NSI}
\label{sec:nsi}

Here, we derive the approximate analytical expressions 
for the fundamental oscillation parameters in matter 
considering all possible lepton-flavor-conserving and lepton-flavor-violating 
NC-NSI\footnote{The authors in Ref.~\cite{Chatterjee:2015gta} derived 
similar expressions in the context of a particular beyond the Standard Model (BSM)
scenario where they considered the presence of long-range flavor-diagonal NSI 
appearing due to abelian $L_e$-$L_{\mu}$ symmetry. In the present work, we adopt a
model independent approach and introduce all possible NSI parameters at-a-time
in the framework. It allows us to study the evolution of mass-mixing parameters 
in a more generalized scheme considering all possible NSI parameters which
has rich phenomenological implications in neutrino oscillation.} 
which are real $\ie$, all the phases associated with the non-diagonal elements 
of the NSI matrix are assumed to be zero.

In order to simplify the subsequent calculations, 
we perform our analysis in the CP-conserving 
scenario $\ie$, we take the standard Dirac CP phase 
$\delta_{\mathrm{CP}}$ to be zero. The elements 
of the effective Hamiltonian $H_f$ in Eq.~\ref{eq:Heff1} 
are then given by,
\begin{align}
(H_{f})_{11} & \,=\, \Delta_{31} \, [\alpha s^2_{12}c^2_{13}+s^2_{13}+\hat{A}]\\
(H_{f})_{12} & \,=\, \frac{\Delta_{31}}{2} \, [\sin{2\theta_{13}}s_{23}(1-\alpha s^2_{12})+\alpha\sin{2\theta_{12}}c_{13}c_{23}+2\varepsilon_{e\mu}\hat{A}] \\
(H_{f})_{13} & \,=\, \frac{\Delta_{31}}{2} \, [\sin{2\theta_{13}}c_{23}(1-\alpha s^2_{12})-\alpha\sin{2\theta_{12}}c_{13}s_{23}+2\varepsilon_{e\tau}\hat{A}] \\
(H_{f})_{22} & \,=\, \frac{\Delta_{31}}{2} \, [\alpha c^2_{12}+c^2_{13}+\alpha s^2_{12}s^2_{13}+\cos{2\theta_{23}}(\alpha c^2_{12}-\alpha s^2_{12}s^2_{13}-c^2_{13}) \nonumber \\
	           & -\alpha\sin{2\theta_{12}}s_{13}\sin{2\theta_{23}} +2\beta\hat{A}]\\
(H_{f})_{23} & \,=\, \frac{\Delta_{31}}{2}[\sin{2\theta_{23}}(c^2_{13}-\alpha c^2_{12}+\alpha s^2_{12}s^2_{13})-\alpha\sin{2\theta_{12}}s_{13}\cos{2\theta_{23}}+2\varepsilon_{\mu\tau}\hat{A}] \\
(H_{f})_{33} & \,=\, \frac{\Delta_{31}}{2}[\alpha c^2_{12}+c^2_{13}+\alpha s^2_{12}s^2_{13}+\cos{2\theta_{23}}(c^2_{13}-\alpha c^2_{12}+\alpha s^2_{12}s^2_{13}) \nonumber \\
	           & +\alpha\sin{2\theta_{12}}s_{13}\sin{2\theta_{23}}+2\gamma\hat{A}]
\end{align}
In the above expressions, we use the abbreviations: 
$\cos\theta_{ij} \rightarrow c_{ij}$, $\sin\theta_{ij} \rightarrow s_{ij}$,
and retain the terms of all orders in $\sin\theta_{13}$ and $\alpha$
which are quite essential in light of the large value of $\theta_{13}$.
To find the effective mixing angles and mass-squared differences 
in the presence of Earth matter potential ($V_{CC}$) and all possible 
NC-NSI parameters, we need to diagonalize the effective Hamiltonian 
$H_{f}$ in Eq.~\ref{eq:Heff1}. We approximately diagonalize $H_{f}$ 
by applying three successive rotations $R_{23} (\theta^m_{23}), 
R_{13} (\theta^m_{13})$, and $R_{12} (\theta^m_{12})$, where
$R_{ij}(\theta^m_{ij})$ is the rotation matrix for the $(i,j)$ block
with the rotation angle $\theta^m_{ij}$. The product of these 
rotation matrices construct a $3 \times 3$ unitary matrix 
\begin{equation}
\tilde{U} \equiv R_{23}\left(\theta_{23}^m\right)R_{13}\left(\theta_{13}^m\right)R_{12}\left(\theta_{12}^m\right) \,,
\label{eq:new-U}
\end{equation}
such that it can almost diagonalize $H_f$
\begin{equation}
\tilde{U}^{T}H_{f}\tilde{U} \simeq \mathrm{Diag} \left(m_{1,m}^2/2E, \, m_{2,m}^2/2E, \, m_{3,m}^2/2E\right) \,,
\label{eq:diagonal-matrix}
\end{equation}
where, the off-diagonal terms after the final rotation are 
quite small ($\sim 10^{-8}$) and can be safely neglected.

Below, we give the expressions for the mixing angles in
matter that we derive by equating the small off-diagonal 
elements to zero after each rotation during the 
diagonalization process:
\begin{equation}
\tan 2\theta^m_{23}\simeq\frac{(c^2_{13}-\alpha c^2_{12}+\alpha s^2_{12}s^2_{13})\sin2\theta_{23}-\alpha\sin2\theta_{12}s_{13}\cos2\theta_{23}+2\varepsilon_{\mu\tau}\hat{A}}{(c^2_{13}-\alpha c^2_{12}+\alpha s^2_{12}s^2_{13})\cos2\theta_{23}+\alpha s_{13}\sin 2\theta_{12}\sin2\theta_{23}+(\gamma-\beta)\hat{A}} \;,
\label{eq:th23m}
\end{equation}
\begin{equation}
\tan2\theta^m_{13}\simeq\frac{\sin2\theta_{13}(1-\alpha {s_{12}}^2)\cos\Delta\theta_{23}-\alpha\sin 2\theta_{12}c_{13}\sin\Delta \theta_{23}+2(\varepsilon_{e\mu}s^m_{23}+\varepsilon_{e\tau}c^m_{23})\hat{A}}{(\lambda_3-\hat{A}-\alpha s^2_{12}c^2_{13}-s^2_{13})} \;,
\label{eq:th13m}
\end{equation}
\begin{equation}
{\tan2\theta^m_{12}}\simeq
\frac{{c^m_{13}}[\alpha\sin2\theta_{12}c_{13}\cos\Delta\theta_{23}+\sin2\theta_{13}(1-\alpha s_{12}^2)\sin\Delta\theta_{23}+2(\varepsilon_{e\mu}c^m_{23}-\varepsilon_{e\tau}s^m_{23})\hat{A}]}{(\lambda_2-\lambda_1)} \;,
\label{eq:th12m}
\end{equation}
where, $\Delta\theta_{23} \equiv \theta_{23}-\theta^m_{23}$ is the deviation of the modified mixing angle $\theta_{23}$ from its vacuum value.
In the above equations, $\lambda_{1}$, $\lambda_{2}$, and $\lambda_3$ take the following forms:
\begin{align}
\label{eq:lambda_3}
\lambda_3=&\nonumber \frac{1}{2}\Big[c^2_{13}+\alpha c^2_{12}+\alpha s^2_{12}s^2_{13}+(\beta+\gamma)\hat{A} \\
&+\frac{(\gamma-\beta)\hat{A}+\alpha\sin2\theta_{12}s_{13}\sin2\theta_{23}+(c^2_{13}-\alpha c_{12}^2+\alpha s_{12}^2s_{13}^2)\cos2\theta_{23}}{\cos2\theta^m_{23}}\Big] \;,
\end{align}
\begin{align}
\label{eq:lambda_2}
\lambda_2=&\nonumber\frac{1}{2}\Big[\alpha c^2_{12}+c^2_{13}+\alpha s^2_{12}s^2_{13}+(\beta+\gamma)\hat{A} \\
&-\frac{(\gamma-\beta)\hat{A}+\alpha\sin2\theta_{12}s_{13}\sin2\theta_{23}+(c^2_{13}-\alpha c_{12}^2+\alpha s_{12}^2s_{13}^2)\cos2\theta_{23}}{\cos2\theta^m_{23}}\Big] \;,
\end{align}
\begin{align}
\label{eq:lambda_1}
\lambda_1=\frac{1}{2}\Big[\lambda_3+\hat{A}+s^2_{13}+\alpha s^2_{12}c^2_{13}
-\frac{\lambda_{3}-\hat{A}-s^2_{13}-\alpha s^2_{12}c^2_{13}}{\cos2\theta^m_{13}}\Big] \;.
\end{align}
Note that throughout the entire paper, 
we consider the propagation of neutrinos 
inside the Earth and assume 
normal mass ordering\footnote{There are two possible patterns of neutrino masses: 
a) $m_3 > m_2 > m_1$, called normal mass ordering (NMO) where 
$\Delta m^2_{31} > 0$ and b) $m_2 > m_1 > m_3$, called inverted 
mass ordering (IMO) where $\Delta m^2_{31} < 0$.} (NMO).
In case of antineutrino propagation, one has to reverse the sign of $V_{CC}$ 
in the above equations which in turn reverses the sign of $\hat{A}$. 
Similarly, to get the corresponding expressions for the inverted mass ordering (IMO),
one has to flip the sign of $\alpha$ as well as the sign of $\hat{A}$ 
in Eqs.~\ref{eq:th23m} to \ref{eq:lambda_1}.

\section{Evolution of Mixing Angles in the presence of NSI}
\label{sec:angle_running}

\begin{table}
\centering
\begin{tabular}{|c|c|c|c|c|c|}
\hline\hline
$\theta_{23}$&$\theta_{13}$&$\theta_{12}$&$\delta_{\mathrm{CP}}$&$\Delta m^2_{21} [\mathrm{eV^2}]$&$\Delta m^2_{31} [\mathrm{eV^2}]$ \\
\hline
$40^{\circ}$, 	$45^{\circ}$, 	$50^{\circ}$&$8.5^{\circ}$&$33^{\circ}$&0&$7.5\times 10^{-5}$&$2.44\times 10^{-3}$ \\
\hline\hline
\end{tabular}
\mycaption{The values of the oscillation parameters used in our analysis. 
First column shows three benchmark values of $\theta_{23}$ that we consider
in our study: maximal mixing ($45^{\circ}$), a possible value in the lower
octant ($40^{\circ}$), and an allowed value in the higher octant ($50^{\circ}$). 
The values of the other parameters are consistent with the present best-fit
values as obtained in various global fit 
studies~\cite{Marrone:2021,NuFIT,Esteban:2020cvm,deSalas:2020pgw}.
We assume normal mass ordering (NMO) throughout the paper.}
\label{table:vac}
\end{table}

In the present section, we study in detail how 
the effective mixing angles in matter $\txm$, 
$\tym$, and $\tzm$ (we derive their expressions 
in Sec.~\ref{sec:nsi}) get modified as functions
of energy and baseline in the presence of all
possible NC-NSI. For this study, we consider 
the three-flavor vacuum oscillation parameters
as given in Table~\ref{table:vac}. To show our
results, we consider two benchmark values of 
the NSI parameters: 0.2 and -0.2.
	
\subsection{Running of $\tzm$}
\label{ssec:th23}

\begin{figure}[htb!]
\centering
\includegraphics[scale=0.6]{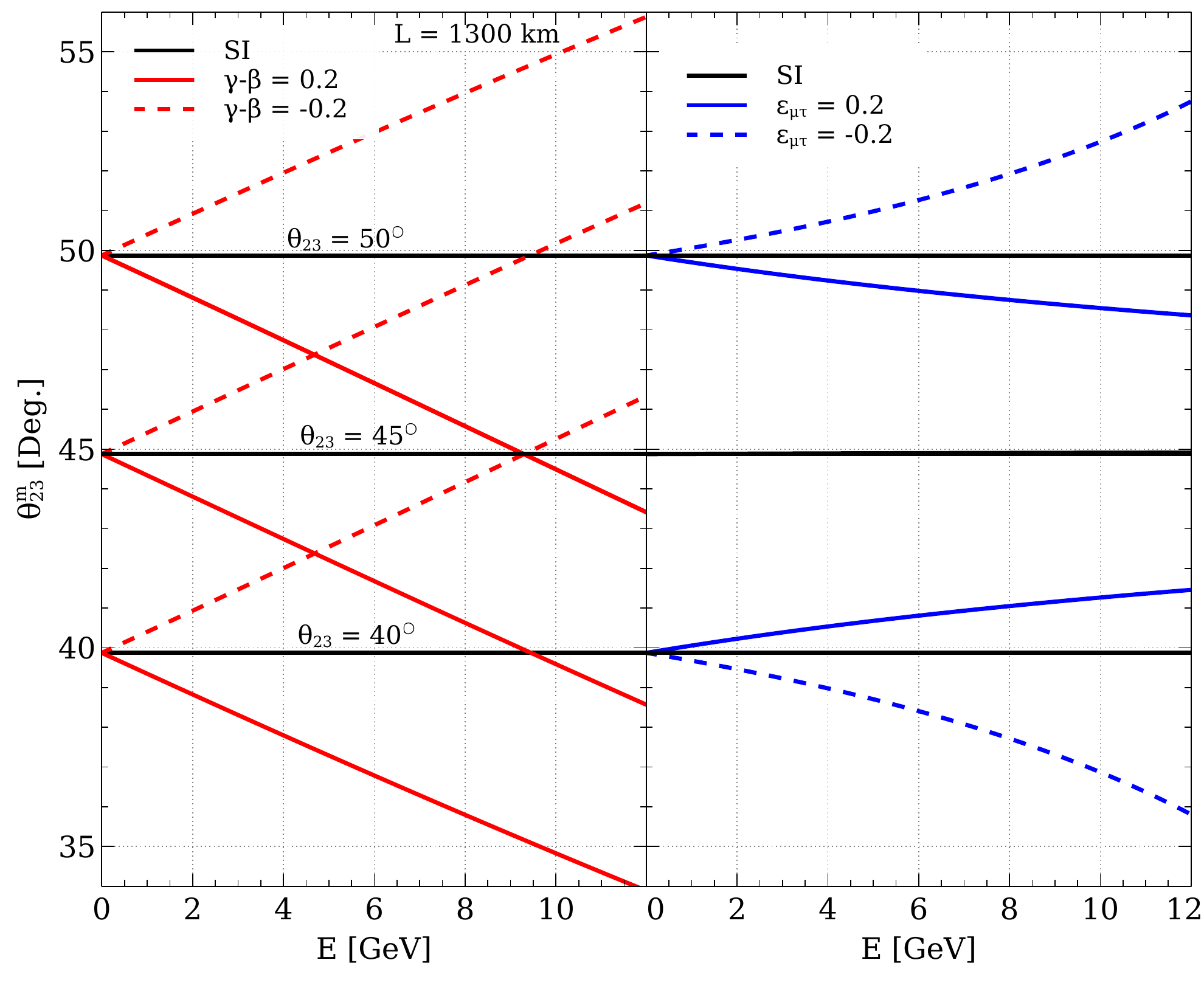}
\mycaption{
Evolution of $\theta^m_{23}$ in matter (given in Eq.\ \ref{eq:th23m}) 
as a function of neutrino energy in the presence of SI and SI+NSI.
Solid black curve in each panel represents the SI case while 
the other curves correspond to the SI+NSI cases 
with positive (solid lines) and negative (dashed lines) 
values of NSI parameters. In the left column, we show 
the running in the presence of NSI parameter 
$(\gamma-\beta)$, while the right column depicts 
the effect of $\emt$. We consider $L$ = 1300 km 
and assume NMO. We present results for three 
different values of $\theta_{23}$ in vacuum:
$40^{\circ}$ (lower octant), $45^{\circ}$ (maximal value), $50^{\circ}$ (upper octant). 
The values of the other oscillation parameters in vacuum are taken 
from Table~\ref{table:vac}.}
\label{fig:th23_1}
\end{figure}

Approximate analytical expression describing the 
evolution of the effective mixing angle $\theta^m_{23}$
is given in Eq.~\ref{eq:th23m}. We can further simplify
this expression by neglecting the small terms which 
are proportional to $\alpha s_{13} \sim 10^{-3}$ in 
Eq.~\ref{eq:th23m}, which enable us to extract the
useful physics insights related to the running of 
$\theta^m_{23}$ in a more concise fashion.
With this approximation, the expression showing 
the evolution of $\theta_{23}$ in matter in the 
presence of NSI takes the form
\begin{align}
\label{eq:th23_1}
\tan 2\theta^m_{23} \simeq \frac{(c^2_{13}-\alpha c^2_{12})\sin2\theta_{23}+2\varepsilon_{\mu\tau}\hat{A}}{(c^2_{13}-\alpha c^2_{12})\cos2\theta_{23}+(\eff) \hat{A}},
\end{align}
where, $\eff = \ett-\emm$.
Two important features
emerge from this simplified
expression.  
\begin{itemize}

\item
Only NSI parameters from the (2,3) block ($\emt$ 
and an effective NSI parameter $\eff \equiv \ett-\emm$) 
of the NSI Hamiltonian contribute to the running 
of $\theta^m_{23}$.

\item
In the limiting case of all NSI parameters equal 
to zero (which in this case removes the standard 
matter effect $\hat{A}$ also), one would get back 
the vacuum mixing angle (\ie, $\tzm = \theta_{23}$)
irrespective of energy, baseline, and the octant 
of $\theta_{23}$. In other words, it implies that 
$\tzm$ does not run in the presence of standard 
matter effect. Note that, in the exact expression 
of $\tzm$ in Eq.~\ref{eq:th23m}, due to the 
presence of the tiny terms proportional to 
$\alpha s_{13}$, $\tzm$ slightly deviates 
from its vacuum value even in the presence 
of SI.

\end{itemize}

In Fig.~\ref{fig:th23_1}, we show the running of 
$\tzm$ (using Eq.\ \ref{eq:th23m}) with energy 
in presence of NSI parameters $(\gamma-\beta)$, $\emt$, 
taken one-at-a-time for a baseline corresponding to the 
DUNE experiment $\ie$ 1300 km.
The left column shows the effect of NSI parameter 
$(\eff)$ while the right column corresponds to the 
effect of $\emt$. The black curves in each column
depict the SI case for three possible values of 
$\theta_{23}$ in vacuum, namely higher octant 
($\theta_{23} = 50^{\circ}$), maximal mixing 
($\theta_{23} = 45^{\circ}$), and lower octant 
($\theta_{23} = 40^{\circ}$). As discussed above, 
value of $\tzm$ in SI case remains almost equal 
to the value of $\theta_{23}$ in vacuum. Only very 
small deviations from the vacuum value of 
$\theta_{23}$ can be observed due to the presence 
of terms proportional to $\alpha s_{13}$ in Eq.~\ref{eq:th23m}, 
which are neglected in Eq.~\ref{eq:th23_1}.
The solid (dashed) red curves in the left column 
of Fig.~\ref{fig:th23_1} illustrate  the presence of $(\eff)$ 
with a benchmark value of 0.2 (-0.2). We observe that 
for all the three values of $\theta_{23}$ mentioned above,
$\tzm$ monotonically decreases (increases) with energy 
when $(\eff)$ is present with a positive (negative) value. 
In the right column, the solid (dashed) blue curves depict 
the case when only $\emt$ is present with a benchmark 
value of 0.2 (-0.2). Interestingly in lower (higher) octant, 
$\tzm$ increases (decreases) for a positive value of $\emt$. 
For maximal mixing, the running of $\tzm$ with energy 
is negligible in the presence of \emt\ and remains almost 
equal to its vacuum value of $45^{\circ}$ 
(since the denominator of Eq.~\ref{eq:th23_1} vanishes). 
The dependence of $\tzm$ running on the choice of 
octant of $\theta_{23}$ in vacuum can be understood 
from the fact that $\cos 2\theta_{23}$ in the denominator
of the R.H.S. of Eq.~\ref{eq:th23_1} changes sign when
$\theta_{23}$ lies in different octants.

\begin{figure}[htb!]
\centering
\includegraphics[height=6cm,width=14cm]{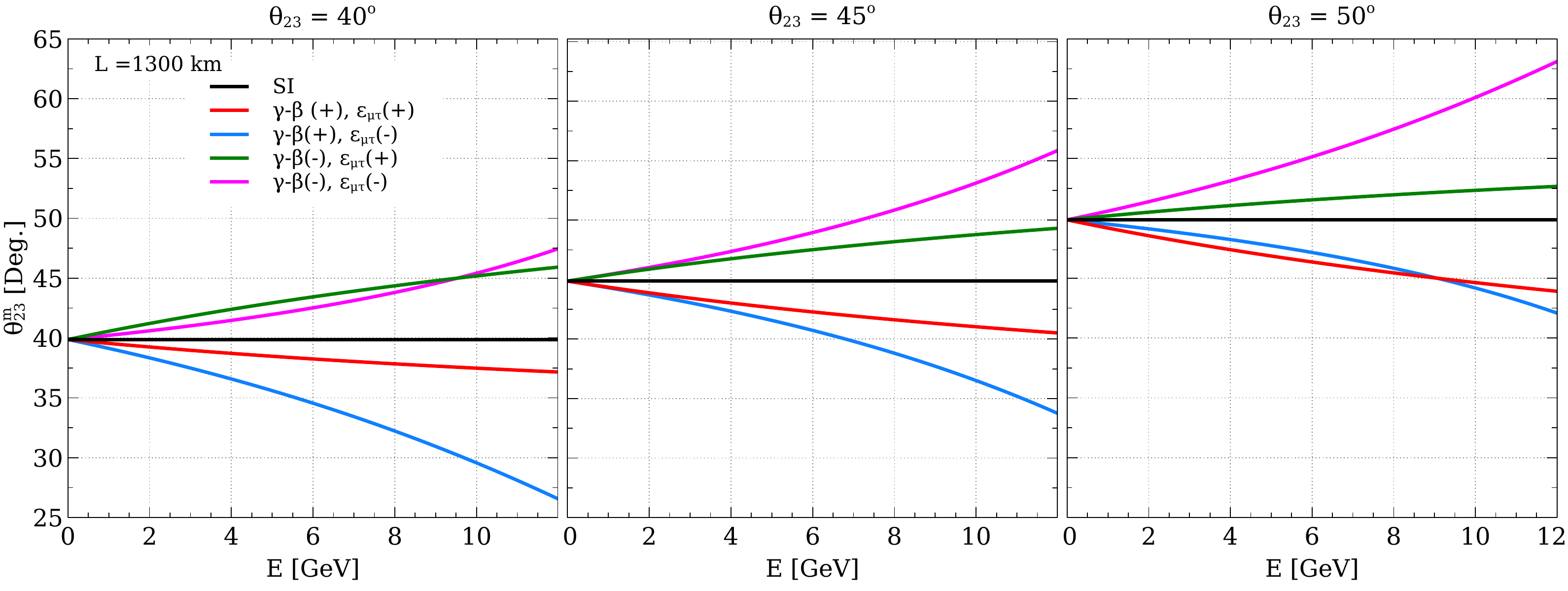}
\mycaption{
Evolution of $\theta^m_{23}$ (given in Eq.~\ref{eq:th23m}) 
with neutrino energy in matter with SI and NSI considering 
both $(\gamma-\beta)$ and $\varepsilon_{\mu\tau}$ 
non-zero at-a-time. Black curve in each column represents 
the SI case while the other curves show the cases with 
four possible combinations of the sign of $(\gamma-\beta)$ 
and $\varepsilon_{\mu\tau}$ with magnitude 0.2. 
The left, middle, and right column correspond 
to the evolution considering three values of 
$\theta_{23}$ in vacuum, $40^{\circ}$, $45^{\circ}$, 
and $50^{\circ}$, respectively. We consider $L$ = 1300 km 
and assume NMO. Values of the oscillation parameters 
in vacuum used in this plot are taken from 
Table~\ref{table:vac}.}
\label{fig:th23_2}
\end{figure}

Fig.~\ref{fig:th23_2} shows the running of $\theta^m_{23}$ 
when both the NSI parameters $\emt$ and $(\eff)$ are non-zero. 
The four colored curves in each panel illustrate the effect 
of the four possible sign combinations of $(\eff)$ and $\emt$ 
while the black curve shows the SI (with standard matter effect and no NSI) case, 
as shown in the legend. As before, three scenarios of the vacuum mixing angle 
$\theta_{23}$ are considered: higher octant ($\theta_{23} = 50^{\circ}$), 
maximal mixing ($\theta_{23} = 45^{\circ}$), and 
lower octant ($\theta_{23} = 40^{\circ}$).    
We note from Fig.\ \ref{fig:th23_2} 
that in the presence of $(\eff)$ with a negative (positive) sign, 
$\tzm$ monotonically increases (decreases) with energy 
irrespective of the sign of $\emt$ and the octant of $\theta_{23}$. 
We also observe that for lower (higher) octant, the decrease (increase) 
is the steepest when $(\eff)$ is positive (negative) with negative 
value of $\emt$. For maximal mixing, the running of $\tzm$ 
appears symmetric around the SI case 
since the term with $\cos2\theta_{23}$ 
in the denominator of Eq.~\ref{eq:th23m} 
vanishes. 

\begin{figure}[htb]
\centering
\includegraphics[scale=0.9]{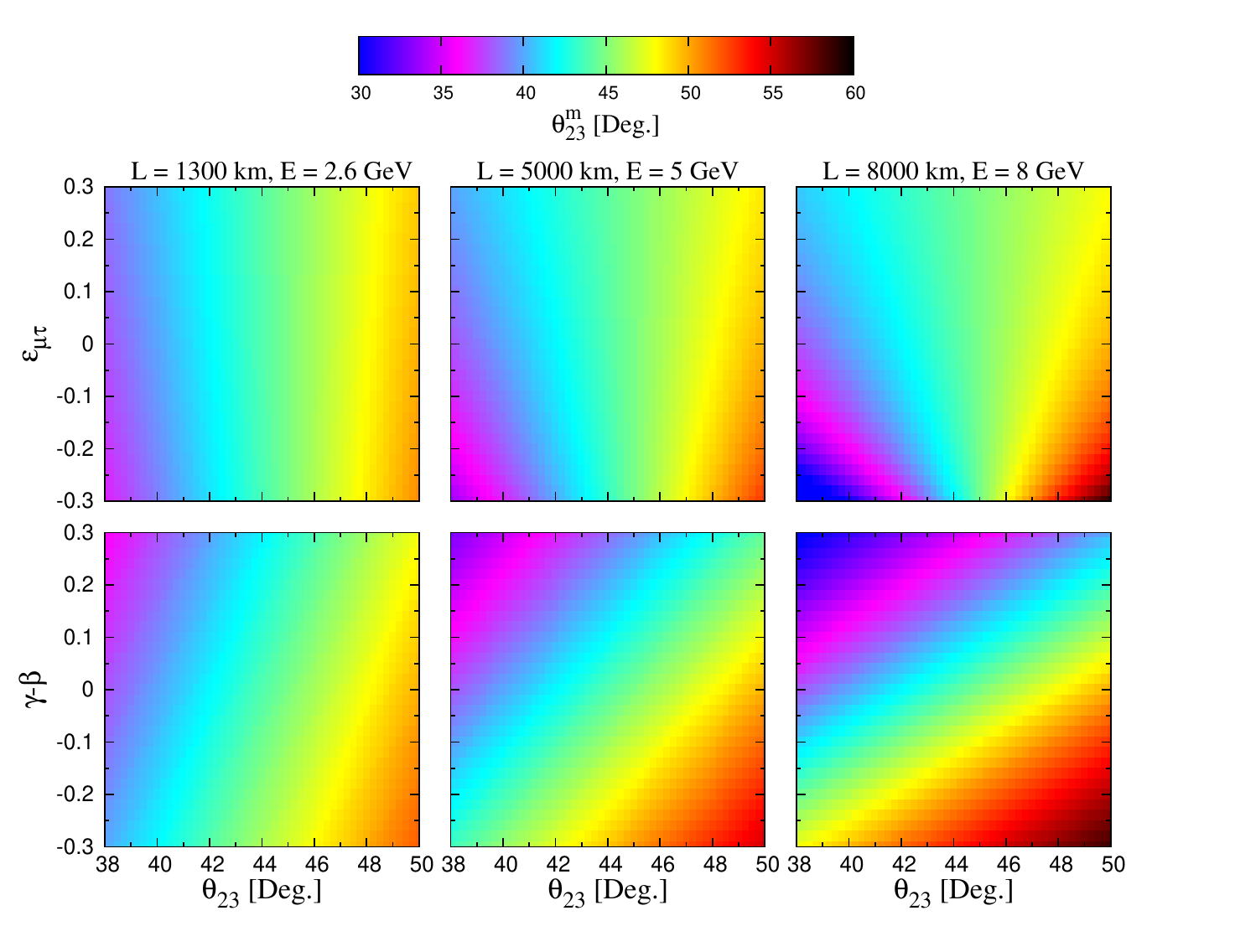}
\mycaption{The evolution of of $\tzm$ 
are shown in the plane of [$\theta_{23}-\emt$] 
(top row) and [$\theta_{23}-(\eff)$] (bottom row). 
The left, middle, and right columns correspond 
to three sets of baseline and neutrino energy, 
namely (1300 km, 2.6 GeV), (5000 km, 5 GeV), 
and (8000 km, 8 GeV), respectively. 
Values of the oscillation parameters 
in vacuum used in this plot are taken 
from Table~\ref{table:vac} and 
we assume NMO.}
\label{fig:th23_3}
\end{figure}

To show a correlation between NSI strength 
and the value of $\theta_{23}$ in vacuum, 
we have shown in Fig.~\ref{fig:th23_3}, 
the evolution of $\tzm$ in the plane of 
[$\theta_{23}-\emt$] (top panels) 
and [$\theta_{23}-(\eff)$] (bottom panels). 
We demonstrate the effect of baseline 
by choosing three different baseline lengths 
as 1300 km, 5000 km, and 8000 km 
in the three columns, respectively.
For each baseline, the energy is chosen 
as the corresponding value near the first 
oscillation maximum ($E_{\text{max}}$) 
for $\nu_{\mu} \to \nu_{e}$ oscillation. 
For the baseline of 1300 km, we see that 
$\tzm$ decreases (increases) from the 
vacuum value ($\theta_{23}$) for a positive 
(negative) $\emt$ at higher octant. However, 
an opposite trend can be observed at lower octant. 
For maximal mixing, $\tzm$ does not change 
in presence of $\emt$ only. 
These features are more pronounced for 
higher baselines since the NSI effect 
(proportional to matter density) gets enhanced. 
In the bottom row, in the presence of positive 
(negative) value of $(\eff)$, $\tzm$ 
decreases (increases) from the vacuum value, 
irrespective of the octant or maximal mixing. 
Larger baselines manifest it more clearly 
as evident from the steeper {\it{slant}} of the boundaries 
between different colors.

As mentioned earlier, we have assumed 
normal mass ordering (NMO) for our analysis. 
In case of inverted mass ordering (IMO) 
with neutrino ($\nu$, IMO), the effect of each 
NSI parameters in $\tzm$ running is reversed 
(\ie, if $\theta^m_{23}$ increases with energy 
in presence of a particular NSI parameter 
with normal ordering of mass, in case of 
inverted mass ordering $\tzm$ will decrease with energy).
This happens since the term $\hat{A}$ associated 
with each NSI parameter changes its sign in case of IMO. 
Also, in case of antineutrino propagation with 
inverted mass ordering ($\bar{\nu}$, IMO), 
running of $\tzm$ is almost the same as that 
of neutrino propagation with NMO ($\nu$, NMO).
This is because of the fact that in both cases, 
sign of $\hat{A}$ is the same.  
	
\subsection{Running of $\tym$}
\label{ssec:th13}
	
Eq.~\ref{eq:th13m} shows the running of 
$\tym$ in matter with NSI. We note that 
all five NSI parameters as well as the 
standard matter effect ($\hat{A}$) have 
impact on the running\footnote{In case of $\tzm$, $\hat{A}$ 
does not affect the running of the parameter.} of $\tym$. 
It is observed that the value of $\theta_{23}$ in vacuum 
(when it is between $40^{\circ}$ and $50^{\circ}$) 
has a very small effect on the running of $\tym$. 
So, we simplify the expression for our understanding 
by assuming that the mixing angle $\theta_{23}$ 
in vacuum is maximal \ie, $45^{\circ}$. 
The relevant expression for the running 
of $\tym$ thus becomes,
\begin{equation}
\label{eq:th13_1}
\tan2\theta^m_{13} \simeq \frac{\sin2\theta_{13}(1-\alpha {s_{12}}^2)(s^m_{23}+c^m_{23})-\alpha\sin 2\theta_{12}c_{13}(c^m_{23}-s^m_{23})+2\sqrt{2}(\varepsilon_{e\mu}s^m_{23}+\varepsilon_{e\tau}c^m_{23})\hat{A}}{\sqrt{2}(\lambda_3-\hat{A}-\alpha s^2_{12}c^2_{13}-s^2_{13})},
\end{equation}
where,
\begin{equation}
\label{eq:lmda3_4}
\lambda_3 = \frac{1}{2}\bigg[{c_{13}}^2+\alpha {c_{12}}^2+\alpha s^2_{12}s^2_{13}+(\beta+\gamma)\hat{A}+\frac{(\gamma-\beta)\hat{A}+\alpha\sin2\theta_{12}s_{13}}{\cos2\theta^m_{23}}\bigg].
\end{equation} 
		    
\begin{figure}[htb!]
\centering
\includegraphics[scale=0.80]{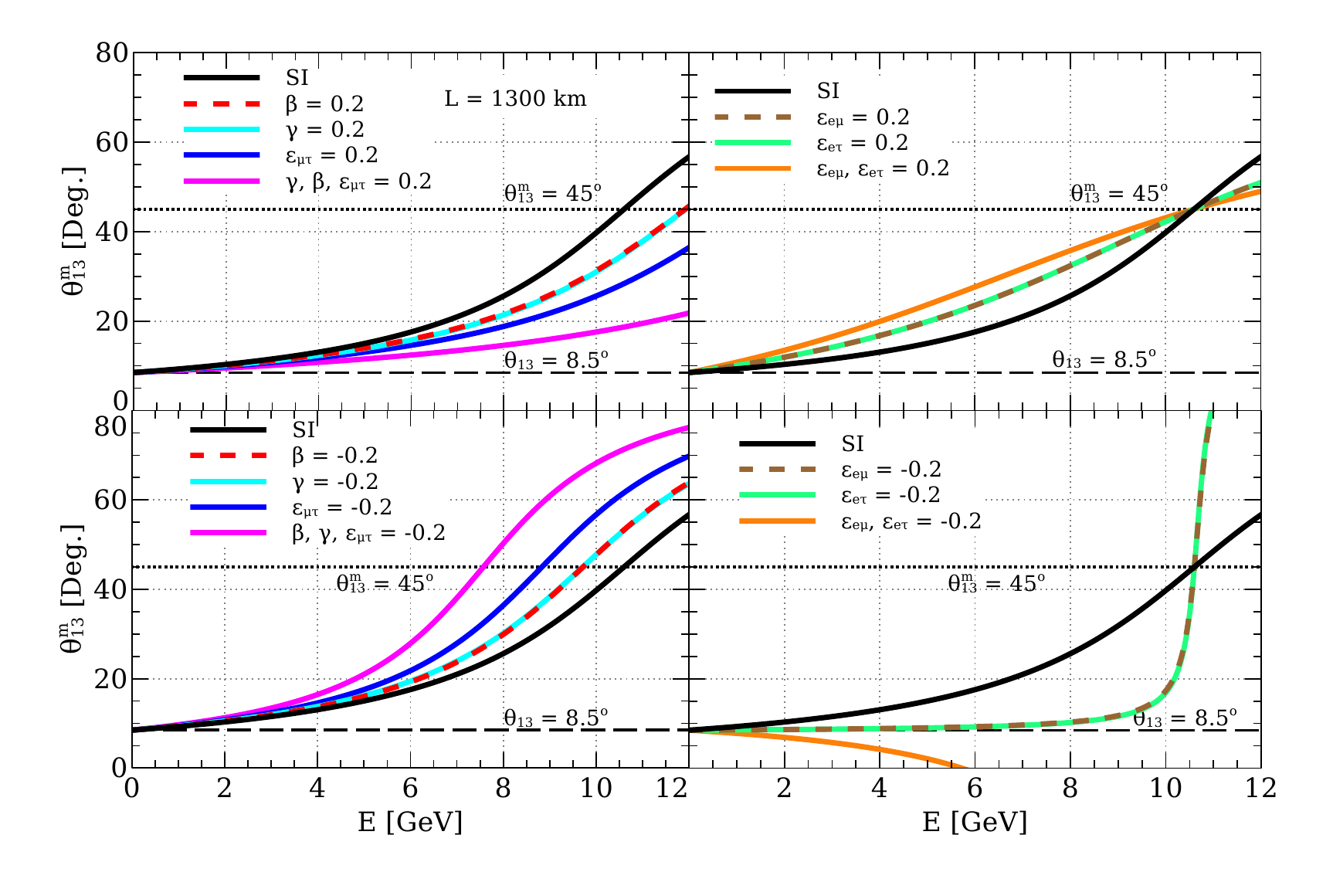}
\mycaption{Evolution of 
$\tym$ (given in Eq.\ \ref{eq:th13_1}) with 
energy in presence of SI and NSI in matter. 
The solid black curve in each panel shows 
the SI case while the other curves correspond 
to the running in presence of SI+NSI. In the 
top (bottom) row, the NSI have been considered 
with a benchmark value of 0.2 (- 0.2). The left 
column depicts the presence of NSI parameters 
in (2,3) block while the right column shows 
the effect of $\eem$ and $\eet$. 
We have used $L$ = 1300 km 
and assumed NMO in the plot. 
Values of oscillation parameters 
in vacuum are given in 
Table~\ref{table:vac} with 
$\theta_{23} =  45^{\circ}$.}
\label{fig:th13_1} 
\end{figure}

In Fig.~\ref{fig:th13_1}, we show the running of $\tym$ 
with energy (by using Eqs.\ \ref{eq:th13_1} and \ref{eq:lmda3_4}) 
in presence of NSI for a baseline of 1300 km and $\theta_{23} = 45^{\circ}$. 
The SI case is depicted by the black curve in each panel 
and the other colored curves indicate the presence 
of NSI parameters in matter with a benchmark strength 
of 0.2 and  -0.2. In the top row, we have shown the 
$\tym$ running when NSI are positive.
The top left panel illustrates the presence of NSI parameters 
in (2,3) block while the right shows the effect of $\eem$ and $\eet$. 
We note that unlike the case of $\tzm$, $\tym$ runs even 
in presence of only SI - its value rapidly rising with energy 
from the vacuum value of $\theta_{13} = 8.5^{\circ}$. 
This can be understood from the fact that with an increase 
in energy, the term ($\lambda_{3}-\hat{A}$) in the denominator 
of the R.H.S. in Eq.~\ref{eq:th13_1} becomes smaller.
The NSI parameters from (2,3) block  suppress the 
rapid rise to some extent due to the modification 
in the value of $\lambda_{3}$ (see Eq.\ \ref{eq:lmda3_4}). 
Moreover, presence of $\beta$ or $\gamma$ only with 
the same strength, makes $\tym$ run in identical 
manner\footnote{From the discussion of Subsec.\ \ref{ssec:th23}, 
we know that $\cos 2\tzm$ is consistently positive (negative with the same magnitude) 
in presence of a positive $\beta$ ($\gamma$) throughout $E > 0$. 
Thus $\lambda_{3}$ in Eq.\ \ref{eq:lmda3_4} remains the same 
in presence of $\beta$ or $\gamma$ with the same strength.}. 
On the other hand, $\eem$ and/or $\eet$ increases the magnitude 
of $\tym$ due to the additional contribution in the numerator 
of the R.H.S. in Eq.\ \ref{eq:th13_1}. 
For the case of maximal mixing of $\theta_{23}$, the impact of 
$\eem$ is identical to that of $\eet$ since $\tzm \simeq \theta_{23} = 45^{\circ}$.  
At lower energy, the gap between the curves showing running 
in presence of $\eem/\eet$ and the SI case increase with energy. 
However, as the value of $\tym$ approaches $45^{\circ}$, 
the gap becomes narrower and at $\tym = 45^{\circ}$, 
these three curves intersect. It happens because, 
around value of $\tym\approx 45^{\circ}$ denominator 
of RHS in Eq.\ \ref{eq:th13_1} becomes so small that 
the effect from the numerator which have $\eem/\eet$ 
is insignificant. In the bottom row, we have shown $\tym$ 
running for the negative values of the NSI parameters. 
It is clear from the bottom left panel that running of 
$\tym$ is enhanced when NSI from (2,3) block is 
present with negative strength. This happens since 
the presence of these negative NSI parameters decrease 
the value of $\lambda_3$, thereby decreasing the overall 
value of the denominator of R.H.S in Eq.~\ref{eq:th13_1}. 
In the bottom right panel, some non-trivial effects are observed. 
We see that negative $\eem$ or $\eet$ highly suppresses 
the running of $\tym$ such that at lower energy ($E$ $\lesssim$ 6 GeV), 
it is almost constant when only one of them is present. 
It can be explained by the fact that both numerator and 
denominator of R.H.S. in Eq.~\ref{eq:th13_1} decreases 
with energy when $\eem$ and/or $\eet$ are negative, 
such that the overall value of $\tym$ remains almost constant at that energy range.
However, at higher energy ($\sim$ 10 GeV) value of 
the denominator is so small that the overall effect led 
to the rapid increase in the magnitude of $\tym$ with energy. 
As we can see from Eq.~\ref{eq:th13_1}, in presence of 
both $\eem$ and $\eet$ with a negative sign, 
the numerator decreases faster with energy 
compared to the previous case due to the additive 
effect of two NSI parameters. Consequently, value 
of $\tym$ decreases with energy from its vacuum value, 
and becomes negative (at $E$ $\gtrsim$ 6 GeV) 
when the numerator becomes negative.  
		
In the case of IMO, the behavior of $\tym$ in SI 
as well as in SI+NSI case is significantly different 
from the NMO case for neutrino. It can be understood 
from the ($\lambda_3-\hat{A}$) term in the denominator 
of Eq.~\ref{eq:th13_1}. Since $\hat{A}$ changes its sign, 
the denominator increases with energy, consequently 
the value of the $\tym$ decreases from its vacuum value. 
However, in case of antineutrino ($\bar{\nu}$) propagation 
and inverted mass ordering ($\bar{\nu}$, IMO), 
since $\hat{A}$ does not change its sign, 
running of $\tym$ is almost similar 
to neutrino ($\nu$, NMO) case.

\subsection{Running of $\txm$}
\label{ssec:th12}
	   
Similar to the case of $\tym$, value of $\theta_{23}$ 
in vacuum (when it is between $40^{\circ}$ and $50^{\circ}$) 
also has very small impact in the running of $\txm$. 
With the assumption of maximal mixing of $\theta_{23}$, 
the relevant expression for $\txm$ in Eq.~\ref{eq:th12m}
takes the form
\begin{equation}
\label{eq:th12_1}
\tan2\theta^m_{12} \simeq \frac{{c^m_{13}}\big[\alpha\sin2\theta_{12}c_{13}(c^m_{23}+s^m_{23})+\sin2\theta_{13}(1-\alpha s_{12}^2)(c^m_{23}-s^m_{23})+2\sqrt{2}(\varepsilon_{e\mu}c^m_{23}-\varepsilon_{e\tau}s^m_{23})\hat{A}\big]}{\sqrt{2}(\lambda_2-\lambda_1)},
\end{equation}
where,
\begin{align}
&  \lambda_2 = \frac{1}{2}\bigg[\alpha c^2_{12}+c^2_{13}+\alpha s^2_{12}s^2_{13}+(\beta+\gamma)\hat{A}-\frac{(\gamma-\beta)\hat{A}+\alpha\sin2\theta_{12}s_{13}}{\cos2\theta^m_{23}}\bigg], \\
&\lambda_1=\frac{1}{2}\Big[\lambda_3+\hat{A}+s^2_{13}+\alpha s^2_{12}c^2_{13}
-\frac{\lambda_{3}-\hat{A}-s^2_{13}-\alpha s^2_{12}c^2_{13}}{\cos2\theta^m_{13}}\Big].
\end{align}
		
\begin{figure}[htb!]
\centering
\includegraphics[scale=0.80]{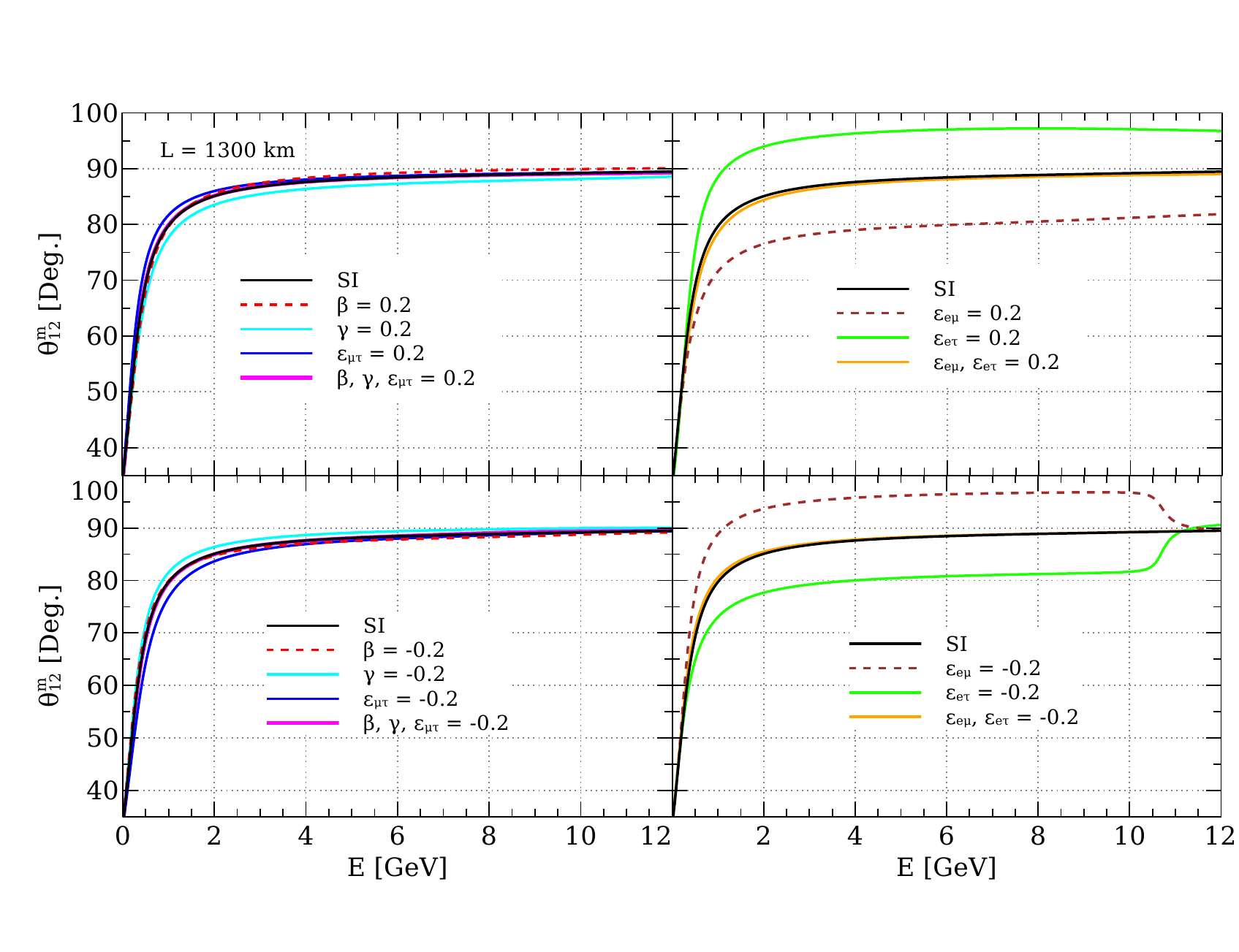} 
\mycaption{Evolution of $\txm$ (given in Eq.~\ref{eq:th12_1}) 
with energy in presence of SI and NSI in matter. 
The solid black curve in each panel shows the SI case 
while the other curves correspond to the running in the 
presence of SI+NSI. In the top (bottom) row, the positive 
(negative) values of the NSI are considered with a 
benchmark value of 0.2 (-0.2). The left column depicts 
the presence of NSI parameters in (2,3) block while 
the right shows the effect of $\eem$ and $\eet$. 
We consider $L$ = 1300 km and assume NMO
to prepare this plot. Values of oscillation parameters 
in vacuum used in this plot are taken from 
Table.~\ref{table:vac} with $\theta_{23} = 45^{\circ}$.}
\label{fig:th12_1}
\end{figure}

In Fig.~\ref{fig:th12_1}, the running of $\txm$ with energy 
is shown both for SI (black curve) and for SI+NSI parameters 
(other curves) for a baseline of 1300 km and $\theta_{23} = 45^{\circ}$. 
The left column shows the effect of the NSI parameters in (2,3) block, 
while the right column depicts the case of $\eem$ and $\eet$ 
with a strength of 0.2 or -0.2. For SI, at small energies 
($E \lesssim 1.5-2$ GeV), $\lambda_{1}$ being close to $\lambda_{2}$,
$\txm$ shows a very steep increase and then quickly saturates 
and approaches to $90^{\circ}$ approximately. 
Saturation occurs due to the following two reasons.
\begin{enumerate}

\item  
With increase in energy, $\lambda_{1}$ 
moves away from $\lambda_{2}$, resulting 
in a large denominator in the R.H.S. of 
Eq.~\ref{eq:th12_1}.

\item  
$\tym$ rises with energy (see Fig.~\ref{fig:th13_1} 
and the relevant discussions in Subsec.~\ref{ssec:th13}) 
and so the overall factor $c_{13}^{m}$ in 
Eq.~\ref{eq:th12_1} decreases.

\end{enumerate}
In the presence of NSI parameters in (2,3) block, 
$\lambda_{1}$, $\lambda_{2}$ and $c^{m}_{13}$ 
undergo mild change, - retaining almost the same 
features as that of SI. 
The presence of $\eem$ ($\eet$) however, 
adds up to the numerator of 
Eq.~\ref{eq:th12_1}\footnote{With increase in energy, 
$\lambda_{2}-\lambda_{1}$ in the denominator of 
Eq.~\ref{eq:th12_1} becomes negative. So a positive 
(negative) contribution to the numerator by $\eem$ 
($\eet$) decreases (increases) the magnitude of $\txm$.} 
and the value of $\txm$ at which it saturates, 
shifts down (up). When both $\eem$ and $\eet$ are present, 
they cancel their effect due to the relative sign between 
them and the running of $\txm$ almost coincides 
with SI scenario. In the bottom row, we have shown 
the running of $\txm$ for the NSI with negative strength.
Since running of $\txm$ very mildly depend on 
NSI parameters from the (2,3) sector (bottom left panel), 
the sign of these NSI parameters do not have any significant 
effect. In the bottom right panel, we see that role of $\eem$ 
and $\eet$ is reversed when the sign of the NSI parameter 
is changed. Interestingly, at energies around 10 GeV, 
sudden decrease (increase) of $\txm$ can be observed 
in the presence of NSI parameter $\eem$ ($\eet$) 
with negative strength. It happens due to the presence 
of the term $\cos\tym$ in the numerator of the R.H.S. of 
Eq.~\ref{eq:th12_1}, which reduces rapidly to a very small 
value around that energy (see Fig.~\ref{fig:th13_1} 
and related discussion in Subsec.~\ref{ssec:th13}).
		
Unlike $\tym$, the $\txm$ running shows similar behavior 
in SI as well as in SI+NSI cases for neutrino propagation 
with IMO ($\nu$, IMO). Also, it shows completely different 
behavior in case of antineutrino propagation with 
inverted mass ordering ($\bar{\nu}$, IMO). 
It can be understood from the fact that in case 
of IMO, sign of first and third terms in the 
numerator of Eq.~\ref{eq:th12_1} gets flipped, 
and in the denominator, the sign of $\lambda_1$ 
gets changed. Since the effect from other remaining 
terms are very small, both numerator and denominator 
change their sign, and as a result, $\txm$ remains 
the same as in the case of ($\nu$, NMO). 
In case of ($\bar{\nu}$, IMO), only first term in the 
numerator changes its sign, $\lambda_1$ in the 
denominator remains the same as in case of 
($\nu$, NMO). As a result, we see a completely 
different behavior of $\txm$. 
		
\section{Evolution of Mass-Squared Differences in the presence of NSI}
\label{sec:mass_running}
		   
After the diagonalization of the effective propagation Hamiltonian 
$H_{f}$ in Sec.~\ref{sec:basics}, we obtain the expressions for the
eigenvalues $m^2_{i,m}/2E$ ($i = 1,2,3$):
\begin{align}
\frac{m^2_{3,m}}{2E} \simeq &\frac{\Delta_{31}}{2}\Big[\lambda_{3}+\hat{A}+s^2_{13}+\alpha s^2_{12}c^2_{13}+\frac{\lambda_{3}-\hat{A}-s^2_{13}-\alpha s^2_{12}c^2_{13}}{\cos2\theta^m_{13}}\Big] \label{eq:m3} \,, \\
\frac{m^2_{2,m}}{2E} \simeq &\frac{\Delta_{31}}{2}\Big[\lambda_1+\lambda_2-\frac{\lambda_1-\lambda_2}{\cos2\theta^m_{12}}\Big] \label{eq:m2} \,, \\
\frac{m^2_{1,m}}{2E} \simeq &\frac{\Delta_{31}}{2}\Big[\lambda_1+\lambda_2+\frac{\lambda_1-\lambda_2}{\cos2\theta^m_{12}}\Big] \,,
\label{eq:m1}
\end{align}
where, we assume $\theta_{23} = 45^\circ$ and 
we are already familiar with the expressions of 
$\theta^{m}_{ij}$ and $\lambda_{i}$. Using the
above equations, we can obtain the approximate 
analytical expressions for the modified mass-squared
differences $\ldmm \equiv m^2_{3,m} - m^2_{1,m}$ 
and $\sdmm \equiv m^2_{2,m} - m^2_{1,m}$. 
The behavior of $\ldmm$ ($\sdmm$) is mainly 
governed by $m^2_{3,m}$ ($m^2_{2,m}$). 
This is due to the fact that in the approximation 
$\txm$ saturating to $90^{\circ}$ (see Subsec.~\ref{ssec:th12}), 
$m^2_{2,m} \approx \lambda_{1} \Delta_{31}$ and 
$m^2_{1,m} \approx \lambda_{2} \Delta_{31}$. 
Therefore, $\lambda_{2}$ being very small, 
$m^2_{1,m}$ is insignificant.

\begin{figure}[htb!]
\centering
\includegraphics[scale=0.80]{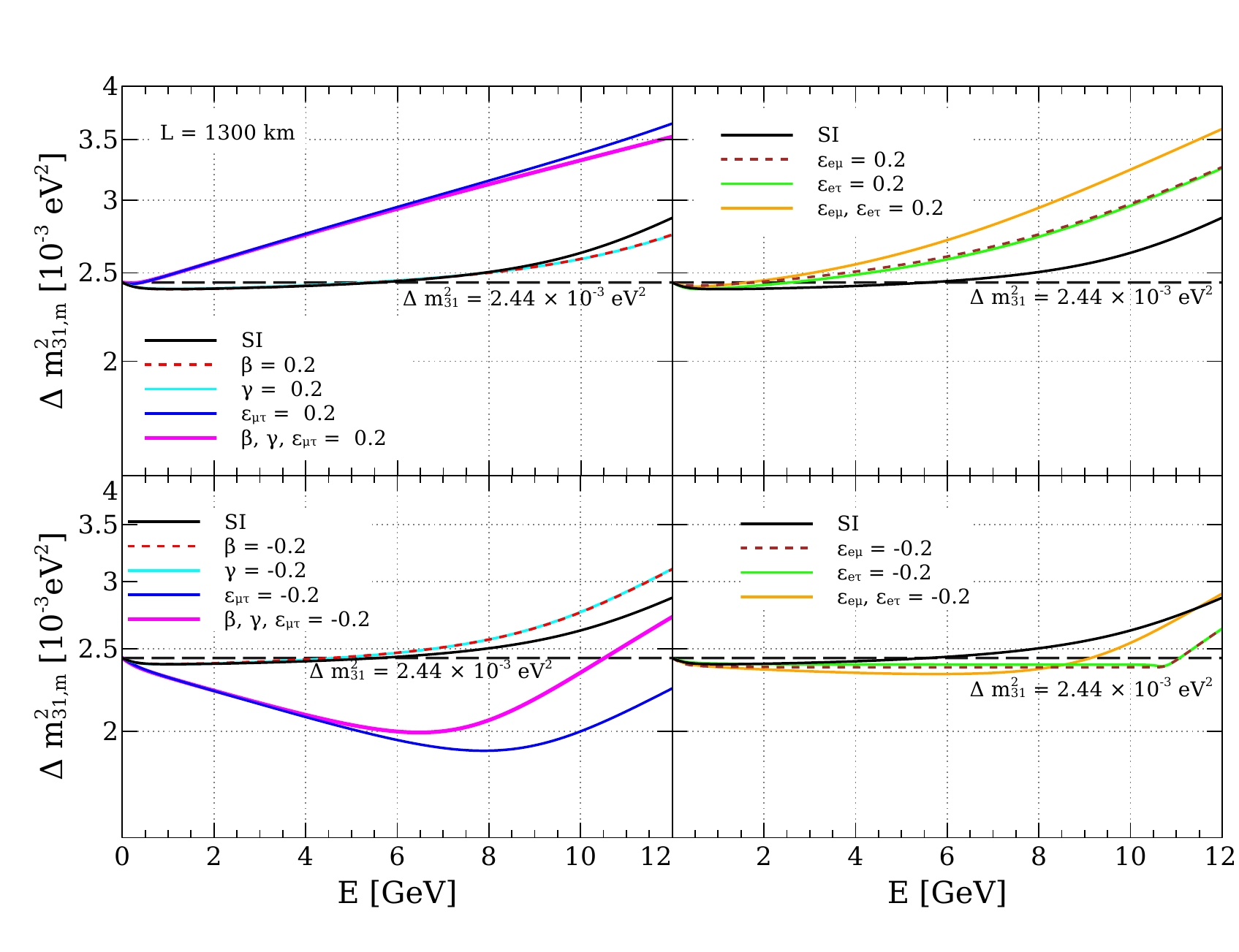}
\mycaption{Variation of $\ldmm$ ($\equiv m^2_{3,m}-m^2_{1,m}$) 
as obtained from Eqs.~\ref{eq:m3}-\ref{eq:m1} is shown with energy 
in SI case and SI+NSI cases. Top (bottom) row corresponds 
to the positive (negative) NSI with strength 0.2. The solid black curve 
in each panel shows the SI case while the other curves show the 
running in presence of NSI. The left column depicts the presence of 
various NSI parameters in (2,3) block while the right shows the effect 
of $\eem$ and $\eet$. We consider $L = 1300$ km and the values 
of the oscillation parameters used in this plot are taken from 
Table~\ref{table:vac}. We assume $\theta_{23} = 45^{\circ}$ and NMO.}
\label{fig:delm31}
\end{figure}

In Fig.~\ref{fig:delm31}, we show the running of $\ldmm$ both 
for SI (black curve) and SI+NSI (other colored curves) scenarios. 
The top (bottom) row corresponds to the running in the 
presence of positive (negative) NSI with strength 0.2. 
The left column depicts the presence of various NSI parameters 
in (2,3) block while the right shows the effect of $\eem$ and $\eet$.   
A baseline of 1300 km and a maximal mixing for $\theta_{23}$ is considered.  
For the SI case, $\ldmm$ first increases very slowly with energy and 
then with a relatively steeper rate (around $E \gtrsim 9$ GeV). 
This is due to the additive contribution of the last term in 
Eq.~\ref{eq:m3} when $\tym$ increases rapidly with energy. 
In the top left panel, presence of $\beta$ or $\gamma$ 
shows a similar running of $\ldmm$ while the introduction of $\emt$ 
shows a steady and almost linear increase with energy due to the 
increase of $\lambda_{3}$ appearing in R.H.S. of Eq.~\ref{eq:m3}. 
In the top right panel, the presence of $\eem$ or $\eet$ shows identical 
effects and makes $\ldmm$ rise with a steeper rate. 
Both $\eem$ and $\eet$ when present together generate 
an additive effect and further elevates the steepness of $\ldmm$. 
In the bottom row, we show the running in the presence 
of negative NSI with strength 0.2. Presence of $\beta$ 
or $\gamma$ with flipped signs reverse the behavior 
of $\ldmm$. In presence of negative $\emt$, initially, 
there is a steady decrease in the value of $\ldmm$ 
because of the decreasing behavior of $\lambda_3$. 
However, at higher energy ($E\gtrsim$ 7 GeV), 
we see sudden growth in the running of $\ldmm$ 
due to increase in the value of $\tym$ at a faster 
rate which in turn increase the value of $m^2_{3,m}$.
In the bottom right panel, we show the running 
in the presence of $\eem$ and/or $\eet$ 
with negative strength. In the presence of negative 
$\eem$ or $\eet$, the value of $\ldmm$ becomes 
almost constant initially ($E\lesssim 10.5$ GeV), 
which can be understood from the running of 
$\tym$ in the presence of negative NSI 
(bottom right panel of Fig.~\ref{fig:th13_1}) 
and the fact that $\lambda_3$ is constant 
in the presence of $\eem$ or $\eet$. 
At $E\gtrsim$ 10 GeV, a sudden increase 
in the value of $\tym$ leads to the increasing 
behavior of $\ldmm$ around that energy.

\begin{figure}[htb!]
\centering
\includegraphics[scale=0.80]{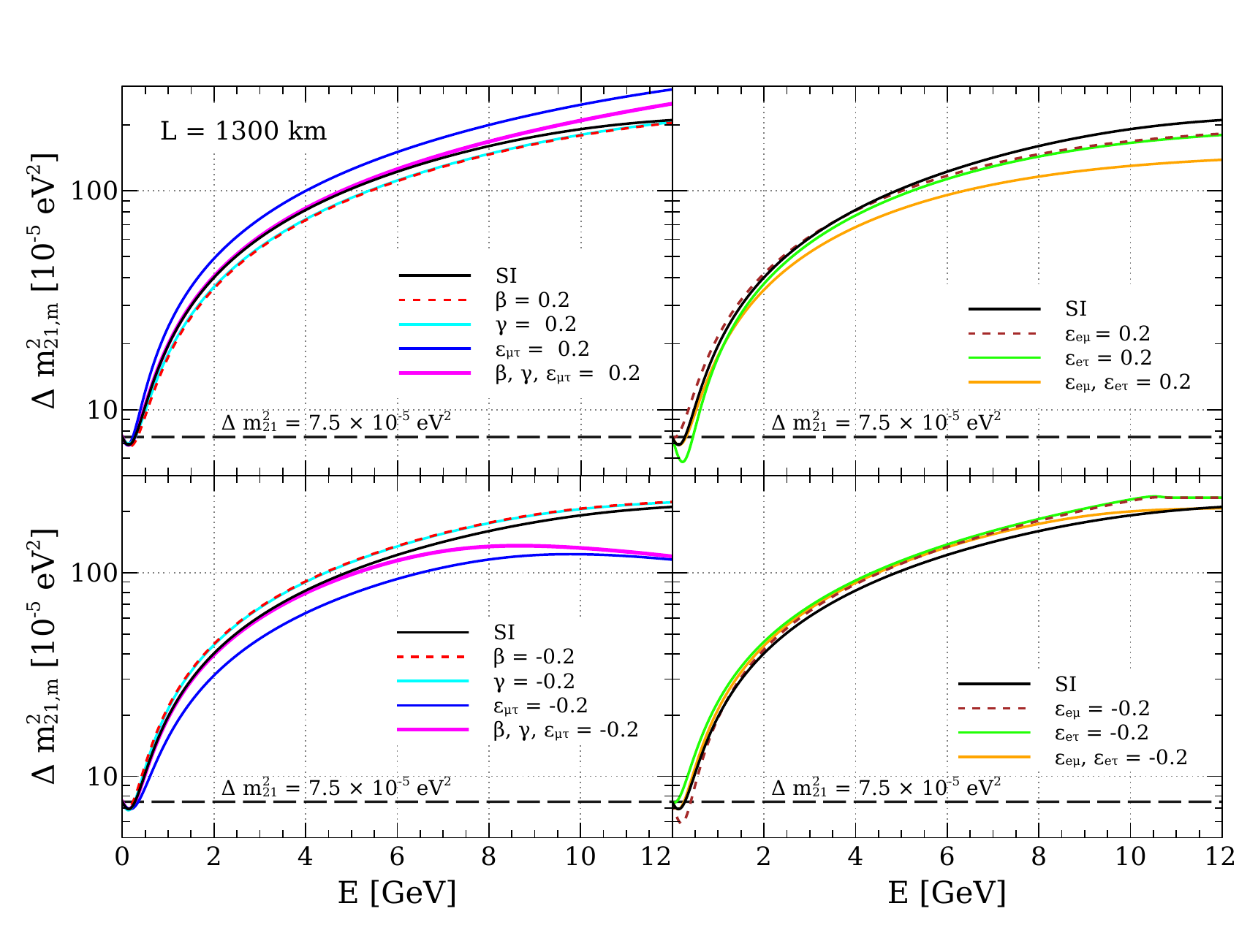}
\mycaption{Variation of $\sdmm$ ($\equiv m^2_{2,m}-m^2_{1,m}$) 
as obtained from Eqs.~\ref{eq:m3}-\ref{eq:m1} is shown with energy 
in SI case and SI+NSI cases. Top (bottom) row corresponds to the 
positive (negative) NSI with strength 0.2. The solid black curve 
in each panel shows the SI case while the other curves show 
the running in presence of NSI. The left column depicts the presence 
of various NSI parameters in (2,3) block while the right shows the 
effect of $\eem$ and $\eet$. We consider $L = 1300$ km and 
the values of the oscillation parameters used in this plot are taken 
from Table~\ref{table:vac}. We assume $\theta_{23} = 45^{\circ}$ and NMO.}
\label{fig:delm21}
\end{figure}

In Fig.~\ref{fig:delm21}, we have shown the running 
of $\sdmm$ with energy for the baseline 1300 km 
and $\theta_{23} = 45^{\circ}$. The black curve 
in each panel corresponds to the SI case while 
other curves show the running in presence of NSI. 
Top (Bottom) row corresponds to the running 
in the presence of positive (negative) NSI with 
strength 0.2. SI case shows steady increase with 
energy, - reaching a value of an order as high as 
$\gtrsim 10^{-3} \text{ eV}^{2}$ from its vacuum 
order of magnitude $10^{-5} \text{ eV}^{2}$. 
In other words, the running of $\sdmm$ can make 
itself comparable in magnitude with that of $\ldmm$. 
In the presence of positive (top row) or negative (bottom row) 
NSI (except for negative $\emt$ or taking negative 
$\beta, \gamma, \emt$ together), the behavior of 
$\sdmm$ does not show significant deviation 
in magnitude from SI case. But interestingly, 
depending on the sign of NSI parameter, 
the magnitude of $\sdmm$ in presence of NSI 
is slightly higher or lower than in presence of SI.  		 
In presence of negative $\emt$ 
(when present singly or together with negative 
$\beta$ and $\gamma$), we see a deviation from 
SI case at higher energy which can be understood 
from the variation of $\lambda_1$ with energy. 
With negative $\eem$ or $\eet$, however, at 
$E\gtrsim$ 10.5 GeV $\sdmm$ becomes almost 
constant. It happens due to a sudden increase 
in the value of $\tym$ around that energy 
which results in saturation of the value 
of $\lambda_1$.
		
In the case of IMO, running of $\sdmm$ is almost 
the same as ($\nu$, NMO) case for both neutrino 
and antineutrino propagation. However, IMO leads 
to a significant change in the running of $\ldmm$ 
for both neutrino and antineutrino propagation 
which is obvious because the vacuum value 
$\ldm$ changes its sign.

\section{$\theta_{13}$-Resonance in the presence of NSI}
\label{sec:resonance}
	 
From the running of $\tym$ (Eq.~\ref{eq:th13_1}), 
we see that interestingly there exists a resonance
such that
\begin{equation}
\label{eq:resonance}
\hat{A} = \lambda_{3}-\alpha s^{2}_{12} c^{2}_{13} - s^{2}_{13} \,.
\end{equation}
Consequently, the denominator of the R.H.S. of Eq.~\ref{eq:th13_1} 
becomes close to zero and $\tym$ becomes maximal ($45^{\circ}$). 
We note that this resonance is independent of the value of $\eem$ 
or $\eet$ (as evident from the right panels of Fig.~\ref{fig:th13_1}) 
but depends upon NSI parameters in the (2,3) block.  
We know that for the SI case, under the one mass scale dominance 
(OMSD) approximation ($\Delta m^2_{31}L/4E>>\Delta m^2_{21}L/4E$), 
the resonance occurs at an energy $E_{\text{res}}$ 
such that~\cite{Akhmedov:2004ny},
\begin{equation}
\label{eq:Eres_SM}
\left[E_{\text{res}}^{\text{SI}}\right]_\text{OMSD} = \frac{\Delta m^2_{31}\cos2\theta_{13}}{2V_{CC}} \,,
\end{equation}
where, $V_{CC}$ is the standard $W$-exchange 
interaction potential in matter (Eq.~\ref{eq:Vcc}).
In presence of NSI, we seek to find out the modifications 
in Eq.~\ref{eq:Eres_SM} considering $\theta_{23} = 45^{\circ}$. 
After replacing $\cos2\theta^m_{23}$ from Eq.~\ref{eq:th23m} 
in the expression for $\lambda_3$ (Eq.\ \ref{eq:lambda_3}), 
we obtain,
\begin{align}
\lambda_3&\simeq\frac{1}{2}\bigg[c^2_{13}+\alpha c^2_{12}+\alpha s^2_{12}s^2_{13}+(\beta+\gamma)\hat{A} \nonumber \\
&+\sqrt{\{\alpha s_{13}\sin2\theta_{12}+(\gamma-\beta)\hat{A}\}^2+\{c^2_{13}-\alpha c^2_{12}+\alpha s^2_{12}s^2_{13}+2\varepsilon_{\mu\tau}\hat{A}\}^2}\bigg] \,.
\end{align}
In the above equation, we neglect the small terms 
proportional to $\alpha s^2_{13}, (\eff)^{2} \hat{A}^{2}$ 
and the cross-term proportional to $\alpha \hat{A} (\eff) s_{13}$. 
Finally, we get the following simpler expression for $\lambda_{3}$, 
\begin{align}  
\label{eq:lmda3_3}    
\lambda_{3} \simeq c^2_{13}+\frac{1}{2}(\beta+\gamma+2\varepsilon_{\mu\tau})\hat{A} \,.
\end{align}
It is noteworthy to mention that for SI case, 
we get $\lambda_{3} \simeq c_{13}^{2}$. 
Putting this back in Eq.~\ref{eq:resonance} 
and using OMSD approximation, we easily obtain 
the well-known expression for resonance in Eq.~\ref{eq:Eres_SM}.
Equating Eqs.~\ref{eq:resonance} and \ref{eq:lmda3_3},
we obtain the following final expression for 
the resonance energy,
\begin{equation}
\label{eq:res}
E_{\text{res}}^{\text{NSI}} \simeq \frac{\Delta m^2_{31}\cos2\theta_{13}}{2V_{CC}} \bigg[\frac{1-(\alpha s^2_{12} c_{13}^{2} /\cos 2\theta_{13})}{1- \frac{1}{2}(\beta+\gamma+2\varepsilon_{\mu\tau})}\bigg] 
= \left[E_{\text{res}}^{\text{SI}}\right]_\text{OMSD} \bigg[\frac{1-(\alpha s^2_{12} c_{13}^{2} /\cos 2\theta_{13})}{1- \frac{1}{2}(\beta+\gamma+2\varepsilon_{\mu\tau})}\bigg] .
\end{equation}
The term in the square bracket in the R.H.S. 
of Eq.~\ref{eq:res} is the correction over 
Eq.~\ref{eq:Eres_SM}. The term 
$\frac{1}{2}(\beta+\gamma+2\varepsilon_{\mu\tau})$ 
is the correction induced by the presence of NSI, 
while $\alpha s^2_{12} c_{13}^{2} /\cos 2\theta_{13}$ 
is the modification induced by relaxing the OMSD approximation. 
Thus it is now also clear analytically that $\tym$-resonance 
gets affected only by the NSI parameters in the (2,3) block 
and not by $\eem$ or $\eet$.

\begin{figure}[htb!]
\centering
\includegraphics[height=12.0 cm, width=14.0 cm]{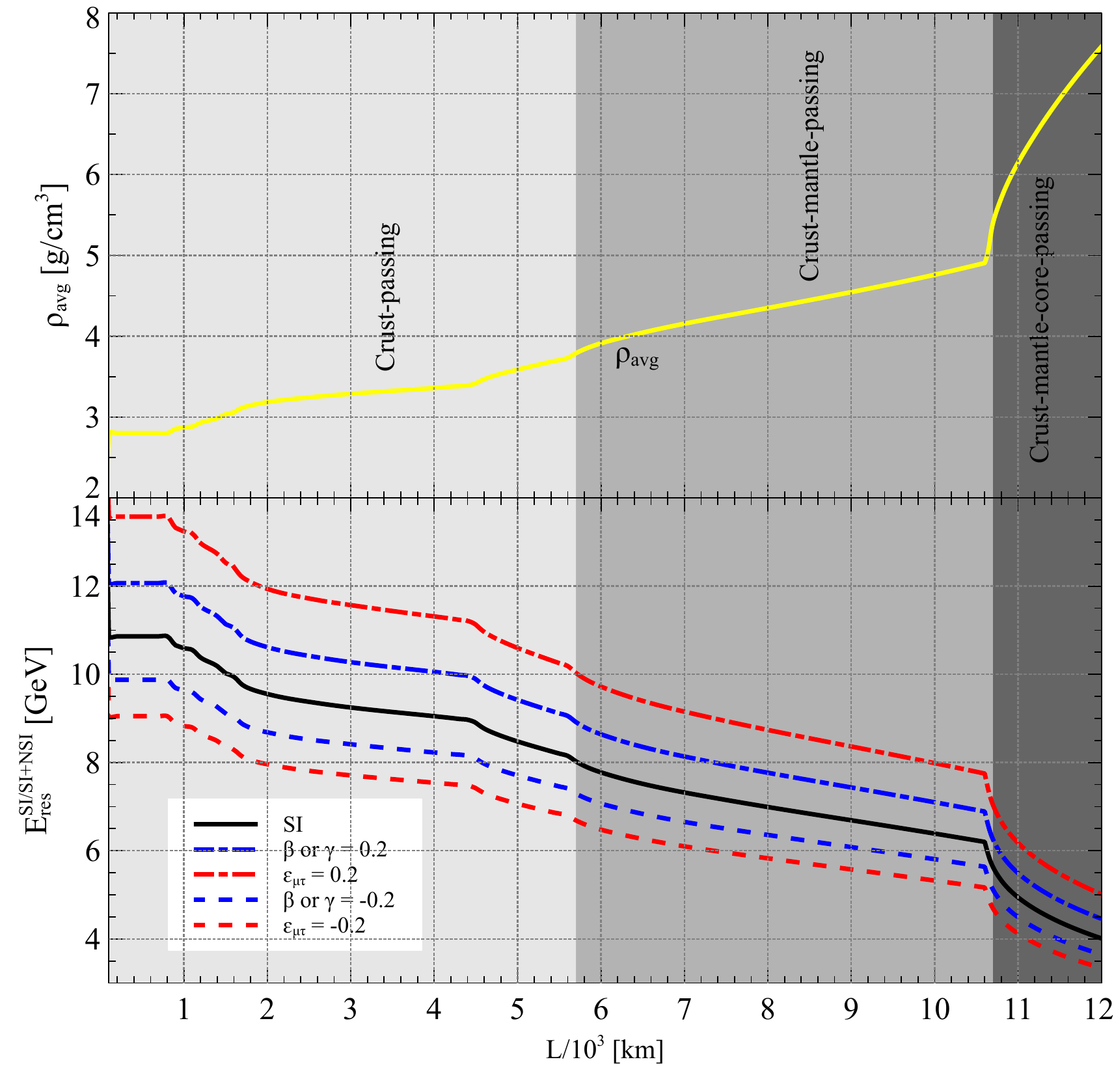}
\mycaption{Behavior of the $\theta_{13}$-resonance 
energy (Eq.~\ref{eq:res}) with baseline length ($L$) 
inside the Earth. Upper panel shows the line-averaged 
constant Earth matter density for a given baseline 
obtained using the PREM profile. Lower panel shows 
the value of the resonance energy ($\tym = 45^{\circ}$) 
corresponding to each baseline length $L$ inside 
the Earth in SI case (solid black line), 
SI + non-zero positive NSI case (dot-dashed lines), 
and SI + non-zero negative NSI case (dashed lines) 
considering one NSI parameters at-a-time as shown
in the legends. The values of the oscillation parameters 
used in this plot are taken from Table~\ref{table:vac}. 
We assume $\theta_{23} = 45^{\circ}$ and NMO.}
\label{fig:res}
\end{figure}

Fig.~\ref{fig:res} (bottom panel) shows 
$E^{\text{NSI}}_{\text{res}}$ as a function 
of baseline length for SI case (black solid curve) 
and in presence of NSI (other colored lines). 
The dot-dashed (dashed) curves depict the case 
of positive (negative) NSI parameters as indicated 
by the legends. The top panel of Fig.~\ref{fig:res} 
shows the line-averaged constant Earth matter density 
($\rho_{\text{avg}}$) for a given baseline $L$ obtained
from the PREM profile~\cite{Dziewonski:1981xy}. 
In both the panels of Fig.~\ref{fig:res}, we indicate 
by three gray shades, the baselines when it touches 
the three interior layers of the Earth: crust, mantle, and core.
Since $\rho_{\text{avg}}$ shows an increase in magnitude 
(thus increasing $V_{CC}$ in Eq.~\ref{eq:res}) with $L$, 
the values of $E_{\text{res}}^{\text{NSI}}$ itself decreases 
with $L$, following similar pattern (for both SI and NSI). 
As it is also clear from Eq.~\ref{eq:res}, a positive (negative) 
value of the NSI parameters $\beta, \gamma$, or $\emt$ 
shifts the magnitude of $E_{\text{res}}^{\text{NSI}}$ 
to a higher (lower) value than the SI case. 
Eq.~\ref{eq:res} also tells us, if it turns out that 
the NSI parameters are present in Nature with such 
magnitudes that $\beta + \gamma = -2\emt$, 
then the correction due to NSI vanishes. In that case, 
if we ignore the minor correction induced by 
$\alpha s^2_{12} c_{13}^{2} /\cos 2\theta_{13}$ term 
we obtain, 
$E_{\text{res}}^{\text{NSI}} \simeq E_{\text{res}}^{\text{SI}}$.

\section{Impact of NSI in $\nu_\mu-\nu_e$ appearance channel}
\label{sec:omsd}

\begin{figure}[htb!]
\centering
\includegraphics[scale=0.9]{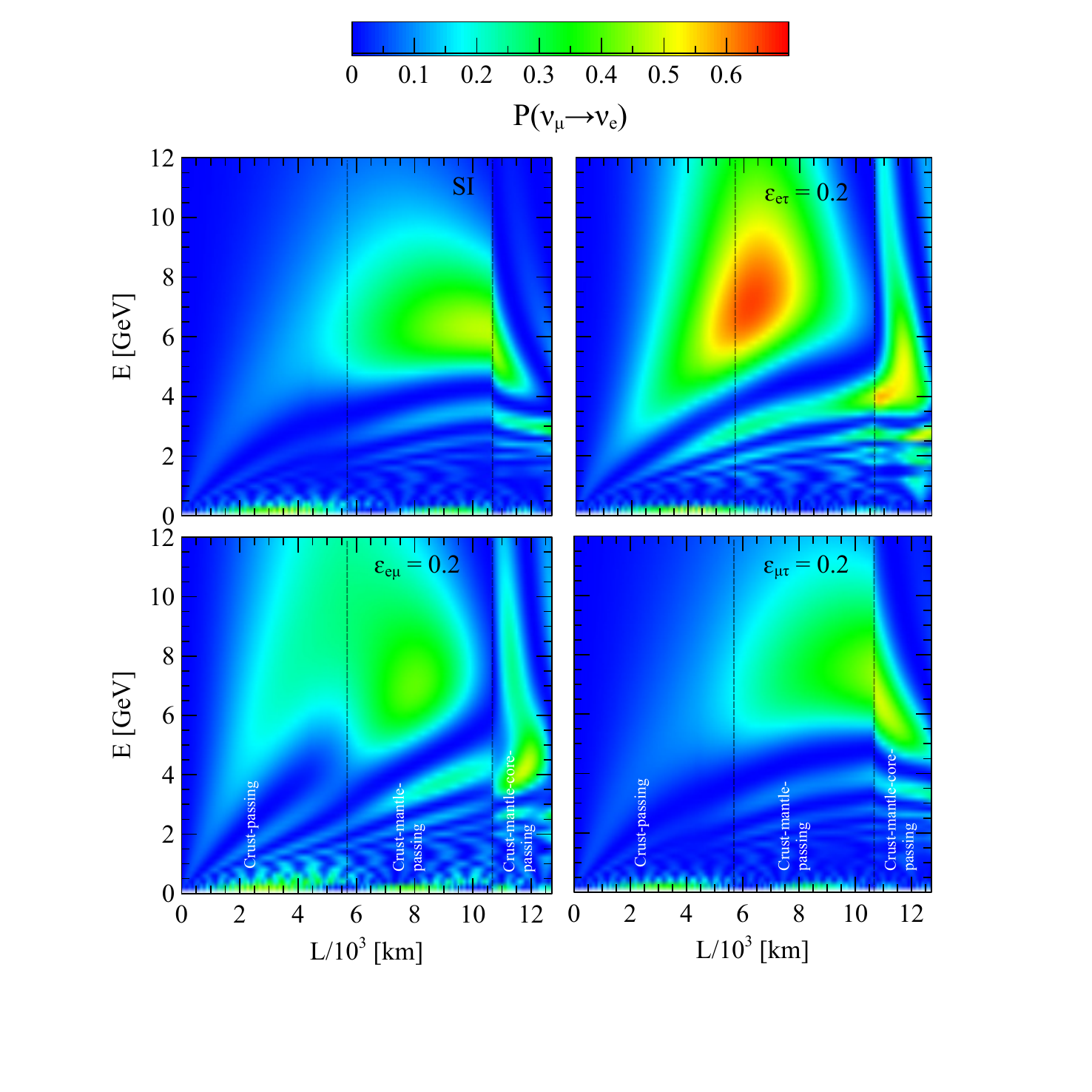}
\vspace*{-1.0cm}
\mycaption{Oscillograms of $\nu_{\mu} \rightarrow \nu_e$ 
transition probability as a function of baseline $L$ and 
energy $E$. Top left panel corresponds to SI case and 
other three panels correspond to the cases in presence 
of non-zero positive NSI parameters (taken one-at-a-time 
with a strength of 0.2 as shown in the legends). 
The values of the oscillation parameters used in this plot 
are taken from Table~\ref{table:vac} with 
$\theta_{23} = 45^{\circ}$ and NMO.}
\label{fig:osc_res_max}
\end{figure}

\begin{figure}[htb!]
\centering
\includegraphics[scale=0.9]{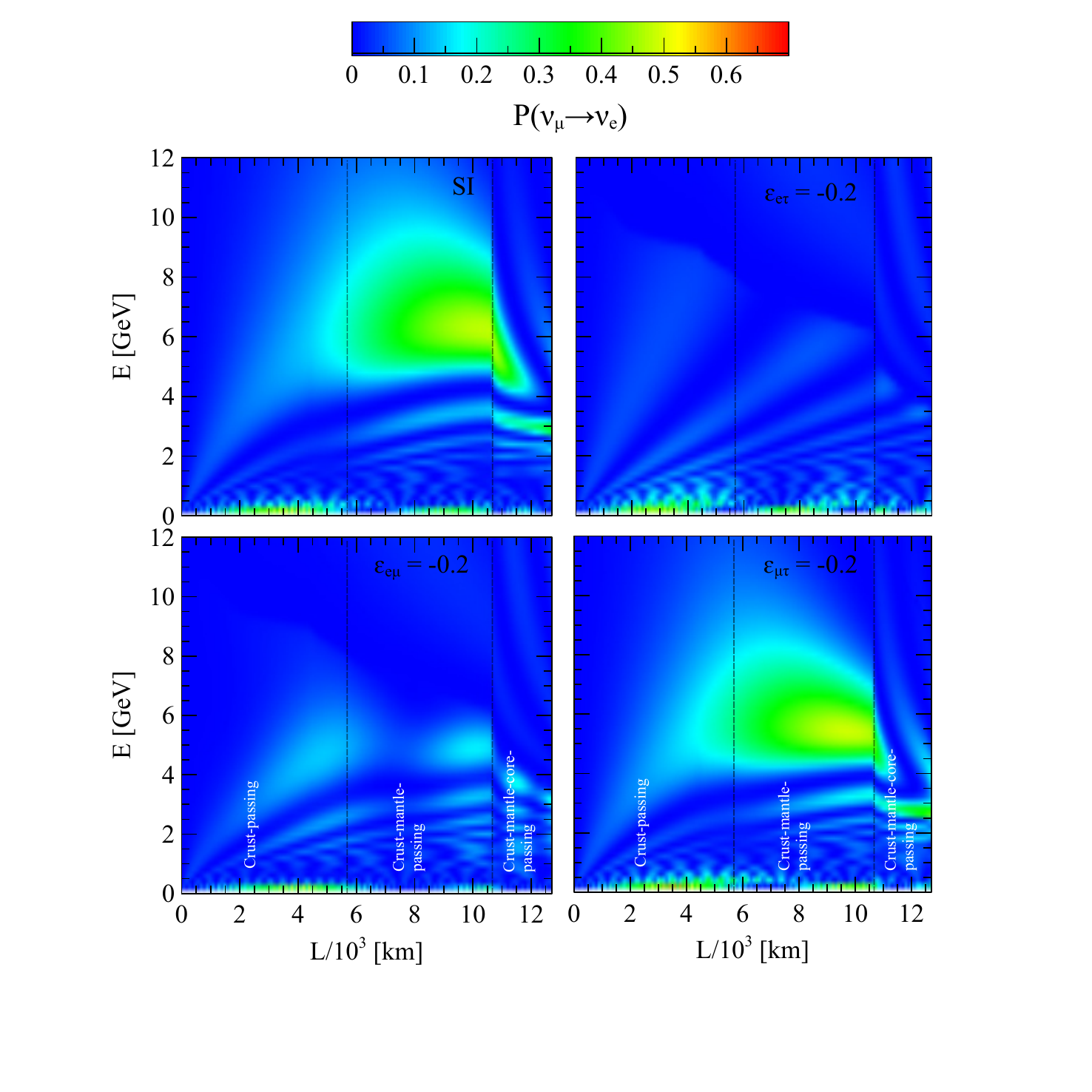}
\vspace*{-1.0cm}
\mycaption{Oscillogram of $\nu_{\mu} \rightarrow \nu_e$ 
transition probability as a function of baseline $L$ and 
energy $E$. Top left panel corresponds to SI case and 
other three panels correspond to the cases in presence 
of non-zero negative NSI parameters (taken one-at-a-time 
with a strength of 0.2 as shown in the legends). 
The values of the oscillation parameters used in this plot 
are taken from Table~\ref{table:vac} with 
$\theta_{23} = 45^{\circ}$ and NMO.}
\label{fig:osc_res_max_neg}
\end{figure}

One of the most important oscillation channel that is probed 
in LBL experiment is $\nu_{\mu}\rightarrow\nu_e$ appearance channel. 
This channel will play a significant role in determining the value of 
the CP phase, neutrino mass ordering, octant of $\theta_{23}$ 
from various upcoming neutrino oscillation experiments. 
So, in this section, we are interested in studying the effect of 
NSI on $\nu_\mu \to \nu_e$ transition probability maxima 
at various baselines ($L$) through the Earth-matter with 
neutrino beam having energy ($E$) in the GeV range. 
In order to study this, in Fig.\ \ref{fig:osc_res_max}, 
we plot the $\nu_{\mu}\rightarrow\nu_{e}$ transition 
probability in $(E$ - $L)$ plane in SI case and 
SI+NSI cases considering a benchmark value of 0.2 
for the strength of the NSI parameters. We use the 
vacuum probability expression for $\nu_{\mu} \to \nu_{e}$ 
appearance~\cite{Agarwalla:2013tza} and replace the vacuum oscillation 
parameters with their modified counterparts in matter with SI and NSI 
(Eqs.~\ref{eq:th23m}-\ref{eq:th12m} and Eqs.~\ref{eq:m3}-\ref{eq:m1}) 
and use this modified probability expression for plotting 
Fig.~\ref{fig:osc_res_max}.  $\theta_{23}$ in vacuum is considered 
to be maximal. We check that Fig.~\ref{fig:osc_res_max} shows 
very good agreement in both SI and SI+NSI cases with the 
exact three-flavor oscillation probabilities which are calculated 
numerically using the GLoBES software~\cite{Huber:2004ka, Huber:2007ji}.
Top left panel shows the SI case where no NSI are taken into account. 
Here, it is observed that the region of maximum appearance probability 
occurs for the baseline almost passing through the core and the mantle boundary. 
However, in presence of $\varepsilon_{e\mu}$ (bottom left panel) 
or $\varepsilon_{e\tau}$ (top right panel) this region shifts towards lower baselines. 
In case of $\varepsilon_{\mu\tau}$ (bottom right panel), this region remains almost 
the same as in the SI case. To show the effect of NSI with negative strength, 
we similarly plot the oscillograms in Fig.~\ref{fig:osc_res_max_neg} 
in SI case and in the presence of negative NSI with strength 0.2. 
Huge differences in the oscillation patterns can be observed 
in case of $\eem$ and $\eet$. Unlike Fig.~\ref{fig:osc_res_max}, 
there is no such region of the maximum transition probability 
in Fig.~\ref{fig:osc_res_max_neg}. 

In Fig.~\ref{fig:osc_res_max}, we notice that for positive values 
of $\eem$ and $\eet$, a significant enhancement in the 
$\nu_{\mu}\rightarrow\nu_{e}$ transition probability as compared 
to SI case, for some choices of $L$ and $E$ where we have large matter effects. 
On the other hand, in Fig.\ \ref{fig:osc_res_max_neg}, for negative choices 
of $\eem$ and $\eet$, we see a large depletion 
in $\nu_{\mu}\rightarrow\nu_{e}$ transition probability 
for some choices of $L$ and $E$ where matter effect is suppressed. 
Now, we make an attempt to understand these features with the help 
of approximate analytical expressions.
After replacing the vacuum oscillation parameters with their modified 
counterparts in the $\nu_{\mu} \to \nu_{e}$ transition probability 
as mentioned above, we simplify it further by using the approximation 
that $\txm$ almost saturates to $\pi/2$ (see Fig.~\ref{fig:th12_1} and 
the related discussion in Subsec.~\ref{ssec:th12}). 
As a result, we obtain the following simplified expression 
that helps us to explain the broad features observed 
in Figs.~\ref{fig:osc_res_max} and \ref{fig:osc_res_max_neg}, 
\begin{equation}
\label{eq:pme2}
P^{m}_{\nu_{\mu}\rightarrow \nu_{e}} = \underbrace{\sin^2\theta_{23}^{m}}_{T_1} \,\, \underbrace{\sin^{2}2\theta^m_{13}}_{T_2} \,\, \underbrace{\sin^{2}\bigg[\frac{1.27\times\Delta m^{2}_{32,m} L}{E}\bigg]}_{T_3} \,.
\end{equation}

\begin{figure}[htb!]
\centering
\includegraphics[scale=0.9]{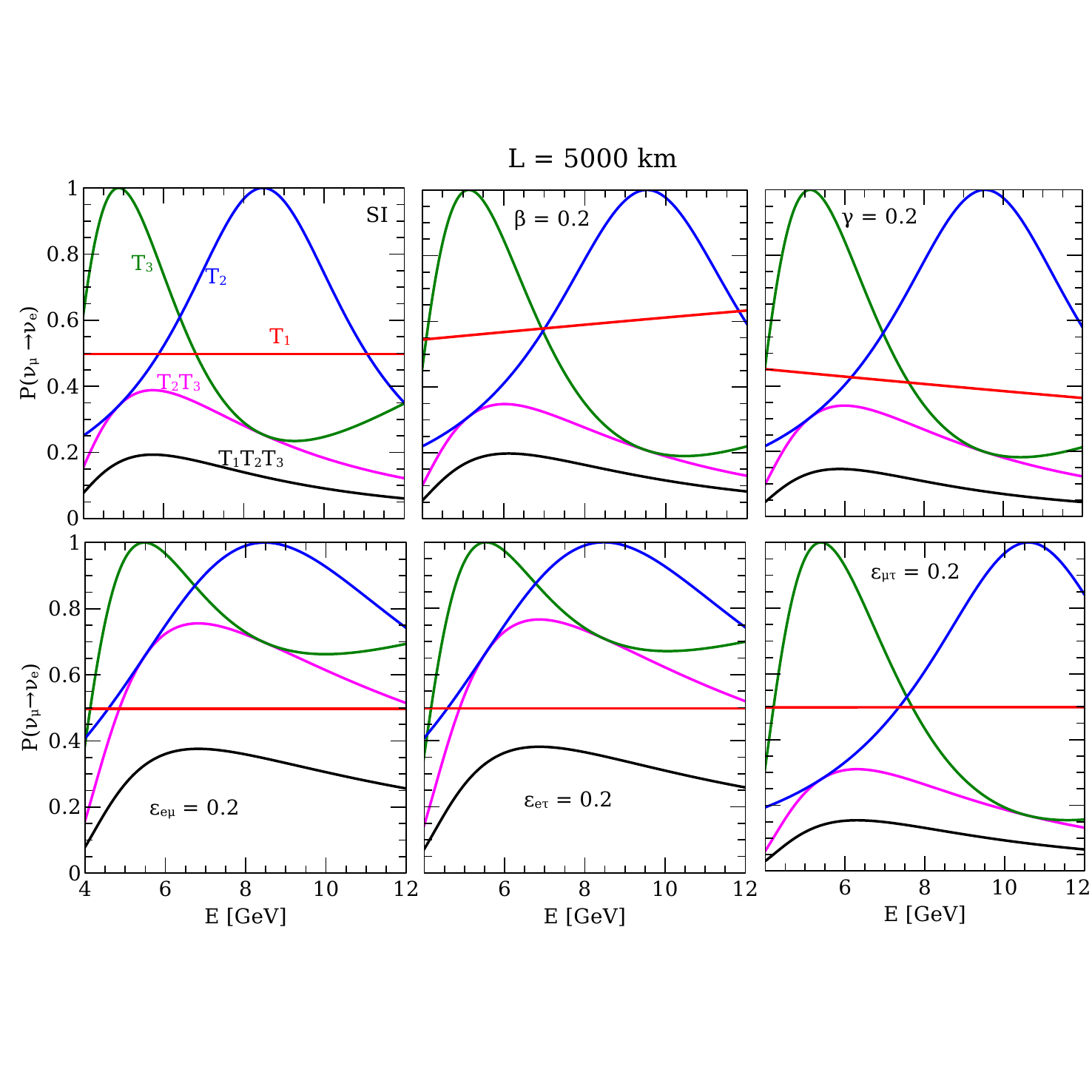}
\vspace*{-1.5cm}
\mycaption{Variation of $\nu_{\mu}\rightarrow\nu_{e}$ 
transition probability (Eq.~\ref{eq:pme2}) with energy 
under the approximation $\txm\rightarrow 90^{\circ}$ 
for a baseline of $L$ = 5000 km. $T_1$ (red curve), 
$T_2$ (blue curve), and $T_3$ (green curve) are the 
three terms defined in Eq.~\ref{eq:pme2}. Various panels 
represent the SI case and SI+NSI cases as shown 
in the labels. To prepare this plot, the values of the
three-flavor oscillation parameters are taken from
Table~\ref{table:vac}. We assume $\theta_{23} = 45^{\circ}$ 
and NMO.}
\label{fig:Pmue_approx}
\end{figure}

In Fig.~\ref{fig:Pmue_approx}, we plot Eq.~\ref{eq:pme2} 
with energy and also the contribution from each term 
$T_1$, $T_2$, and $T_3$, separately. It is clear from the figure 
that the variation of $T_1$ with energy is very less compared 
to the variation of $T_2$ and $T_3$. Thus, the energy at 
which maximum of $T_1T_2T_3$ occurs is the same as that 
of $T_2T_3$. In other words, the maximum of 
$P^{m}_{\nu_{\mu}\rightarrow\nu_{e}}$ occurs at an energy 
determined by $T_2$ and $T_3$, not $T_1$. 
This feature is also valid for any other baselines.
So, $P^{m}_{\nu_{\mu}\rightarrow\nu_{e}}$ is maximum when 
both the following two conditions are satisfied simultaneously.
\begin{itemize}

\item 
$T_2 \equiv \sin^2\theta^m_{13} = 1$ \textit{i.e.},  $\tym = 45^{\circ}$ 
($\theta_{13}$-resonance condition). This condition is achieved in SI case
when $E = E_{\text{res}} =  \frac{\Delta m^2_{31}\cos2\theta_{13}}{2V_{CC}}$ 
(see Eq.\ \ref{eq:Eres_SM}) with the OMSD approximation.

\item 
$T_3\equiv \sin^{2}\bigg[\frac{1.27\times\Delta m^{2}_{32,m} L}{E}\bigg] = 1$ 
for some energy $E=E^m_{\text{max}}$, such that
\begin{equation}
\label{eq:Emax}
E^m_{\text{max}} = \frac{1.27\times\Delta m^{2}_{32,m} L}{(2n+1)\pi/2} \quad \text{ with } n = 0, 1, 2...
\end{equation}

\end{itemize}
Thus, the maximum matter effect is obtained 
when the condition $E_{\text{res}} = E^m_{\text{max}}$ 
is satisfied~\cite{Banuls:2001zn,Gandhi:2004md,Gandhi:2004bj}.

In order to simplify the expression of $E^m_{\text{max}}$ in the
presence of SI only, we use Eq.~\ref{eq:m3} and Eq.~\ref{eq:m2} 
to calculate $\Delta m^2_{32,m}$ considering all the NSI parameters 
to be zero. Applying the OMSD approximation and 
$\theta_{23} = 45^\circ$, we obtain
\begin{equation}
\label{eq:delm32m_SI}
\Delta m^2_{32,m} = \Delta m^2_{31}\sqrt{(\lambda_3-\hat{A}-s^2_{13})+\sin^22\theta_{13}} \,.
\end{equation}
Now, using Eq.~\ref{eq:delm32m_SI} in the expression 
of $E^m_{\text{max}}$ in Eq.~\ref{eq:Emax}, the condition 
for the maximum matter effect $E_{\text{res}} = E^m_{\text{max}}$ 
gets further simplified. Ultimately, we obtain a simple and compact
relation between the baseline ($L$) and the corresponding
line-averaged constant matter density ($\rho_{\text{avg}}$)
to have the maximum $\nu_{\mu}\rightarrow\nu_{e}$ transition 
probability in matter 
\begin{equation}
\big(\rho_{\text{avg}} L \big)_{\text{SI}} = \frac{(2n+1)\times\pi\times5.18\times10^3}{\tan2\theta_{13}} \hspace{0.1cm} \text{km g/cm$^3$} \,.
\label{eq:rho_l}
\end{equation} 
Note that under the OMSD approximation, the resonance 
energy condition in Eq.~\ref{eq:resonance} takes a very
simple form: $(\lambda_3 - \hat{A} -s^2_{13}) = 0$ and 
we make use of this expression in Eq.~\ref{eq:delm32m_SI}
to obtain Eq.~\ref{eq:rho_l}, which exactly matches with the
expression derived by the authors in Ref.~\cite{Gandhi:2004md}.

Now, we analyze how Eq.~\ref{eq:rho_l} gets modified in the
presence of NC-NSI. First, we use Eqs.~\ref{eq:lambda_3}$-$\ref{eq:lambda_1} 
and Eqs.~\ref{eq:m3}$-$\ref{eq:m1} to derive the following two expressions 
for $m^{2}_{3,m}$ and $m^{2}_{2,m}$ under the OMSD approximation
and assuming $\theta_{23} = 45^{\circ}$.
\begin{align}
m^{2}_{3,m} &= \frac{\ldm}{2}\big[ 
\lambda_{3} + \hat{A} + s^{2}_{13} + T
\big] \,, \nonumber \\
m^{2}_{2,m} &= \frac{\ldm}{2}\big[ 
\lambda_{3} + \hat{A} + s^{2}_{13} - T
\big] \,, 
\end{align}
where,
\begin{equation}
T = \frac{1}{\sqrt{2}}\sqrt{2\big[\lambda_{3} - \hat{A} - s^{2}_{13}\big]^{2} 
+ \big[\sin 2\theta_{13}(c^{m}_{23}+s^{m}_{23}) + 2\sqrt{2}(\eem s^{m}_{23} + \eet c^{m}_{23}) \big]^{2}} \,.
\end{equation}
Using the resonance energy condition (Eq.~\ref{eq:resonance}), we now have
\begin{align}
\Delta m^{2}_{32,m} = m^{2}_{3,m} - m^{2}_{2,m} 
= \ldm \big[ \frac{1}{\sqrt{2}}
\sin 2\theta_{13}(c^{m}_{23}+s^{m}_{23}) + 2(\eem s^{m}_{23} + \eet c^{m}_{23})
\big] \,.
\end{align}
Replacing $\Delta m^{2}_{32,m}$ in Eq.~\ref{eq:Emax}, 
we finally have the following condition for the 
maximal $\nu_{\mu} \rightarrow \nu_{e}$ transition probability
in the presence of NC-NSI.
\begin{align}
\label{eq:rho_l_nsi}
&\big(\rho_{\text{avg}} L \big)_{\text{NSI}}
\nonumber\\ 
& \simeq \frac{(2n+1)\pi \times 5.18 \times 10^{3}}{\tan 2\theta_{13}\big[1-\big\{(\beta+\gamma+2\varepsilon_{\mu\tau})(\frac{c^{m}_{23}+s^{m}_{23}}{2\sqrt{2}})\big\} + \big\{2(\eem s^{m}_{23}+\eet c^{m}_{23})/\tan 2\theta_{13}\big\}\big]} \hspace{0.1cm} \text{km g/cm$^3$}\\
&=   \big(\rho_{\text{avg}} L\big)_{\text{SI}} \bigg[ \frac{1}{1-\big\{(\beta+\gamma+2\varepsilon_{\mu\tau})(\frac{c^{m}_{23}+s^{m}_{23}}{2\sqrt{2}})\big\} + \big\{2(\eem s^{m}_{23}+\eet c^{m}_{23})/\tan 2\theta_{13}\big\}}\bigg]\text{km g/cm$^3$}.  
\end{align} 
The second factor in the R.H.S. of Eq.~\ref{eq:rho_l_nsi} 
is the correction introduced by the NSI parameters. 
As shown in Fig.~\ref{fig:Pmue_approx},
the modified $\tzm$ does not run significantly and 
is also close to $45^{\circ}$ for maximal mixing of $\theta_{23}$. 
Using the approximation $\tzm \simeq 45^{\circ}$, 
we simplify Eq.~\ref{eq:rho_l_nsi} to the following.
\begin{align}    
&\big(\rho_{\text{avg}} L \big)^{max}_{\text{NSI}} 
\simeq  \big(\rho_{\text{avg}} L \big)_{\text{SI}}  \bigg[ \frac{1}{1-\big\{(\beta+\gamma+2\varepsilon_{\mu\tau})\big\}/2 + \big\{\sqrt{2}(\eem +\eet )/\tan 2\theta_{13}\big\}}\bigg]\text{km g/cm$^3$} \,.
\label{eq:rho_l_nsi_maximal}        
\end{align}  

\begin{figure}[htb!]
\centering
\includegraphics[height=12.5 cm, width=13.5 cm]{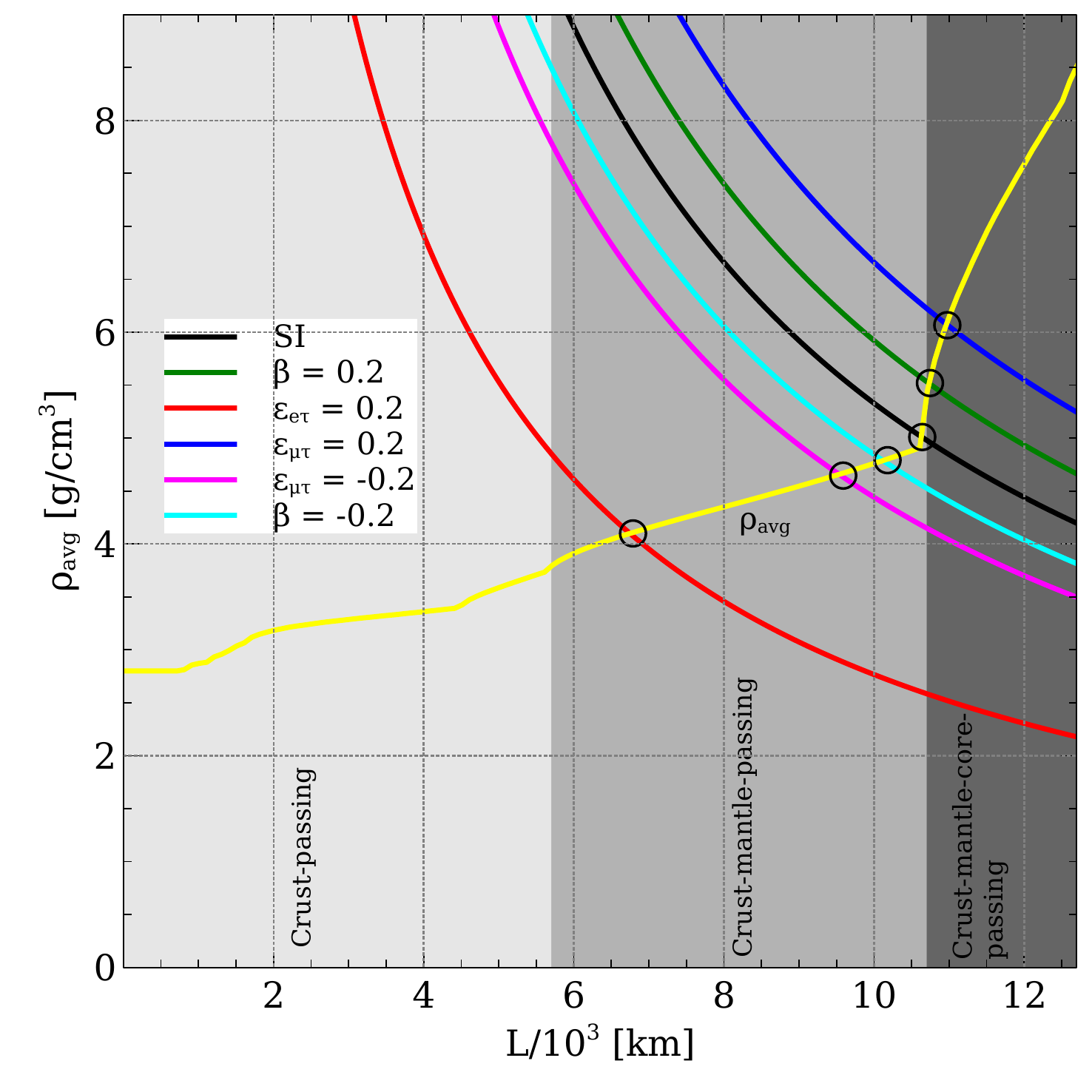}
\mycaption{\label{fig:rhol_cond}The solid yellow line shows 
the line-averaged constant Earth matter density 
(obtained using the PREM profile) for a given $L$. 
The intersection point of the solid black and yellow 
lines depict the value of $L$ and its corresponding $\rho_{avg}$ 
for which $\nu_{\mu}\rightarrow\nu_{e}$ transition probability 
attains the maximum value (see Eq.~\ref{eq:rho_l}). 
The intersection points of the other solid lines with 
solid yellow line show the same considering one 
NSI parameter at-a-time (see the legends). 
The three-flavor oscillation parameters in vacuum 
are taken from Table~\ref{table:vac} with a choice 
of $\theta_{23} = 45^{\circ}$ and NMO.}
\label{fig:osc_max}
\end{figure}

In Fig.~\ref{fig:rhol_cond}, we plot the R.H.S. of 
Eq.~\ref{eq:rho_l_nsi_maximal} in presence of SI 
and also in presence of NSI parameters 
($\beta, \emt, \eet$ taken one-at-a-time with a magnitude of 0.2) 
in the $\rho_{\text{avg}}$ and $L$ plane. In the presence of NSI 
parameter $\gamma$ ($\eem$), Eq.~\ref{eq:rho_l_nsi_maximal} 
is same as in the presence of $\beta$ ($\eet$). 
Again, three gray regions correspond to the baseline length 
passing through the crust, crust-mantle, and crust-mantle-core. 
In the same figure, we also show the line-averaged 
constant Earth matter density according to the PREM profile (yellow curve). 
The points of intersections of the hyperbolic curves 
(corresponding to Eq.~\ref{eq:rho_l} for the SI case 
and Eq.~\ref{eq:rho_l_nsi_maximal} for the SI+NSI cases) 
with the yellow curve give the baseline lengths, required 
to achieve the maximum of $P^{m}_{\nu_{\mu} \to \nu_e}$.
When there are no NSI present in the scenario, 
the baseline length required for maximum transition probability 
is around 10600 km, which almost touches the core of the Earth. 
This same feature is also observed in the oscillogram plot in 
Fig.~\ref{fig:osc_res_max} which is plotted using full 
three-flavor $\nu_{\mu} \rightarrow \nu_e$ oscillation 
probability expression. When NSI parameters from (2,3) block 
with positive (negative) strength are present one-at-a-time, 
the required baseline length increases (decreases) slightly.
But interestingly, when the NSI parameter $\eem$ or $\eet$ 
is present, the required baseline length for maximum $\nu_{\mu} \to \nu_{e}$ 
transition decreases drastically to around 6700 km 
(intersection between red and yellow curve in Fig.~\ref{fig:osc_max}), 
which passes through only crust and mantle of 
the Earth\footnote{From Eq.~\ref{eq:rho_l_nsi}, it is clear that the presence 
of $\beta, \gamma$ or $\emt$ with a positive value increases the NSI correction factor, 
while $\eem$ and $\eet$ decreases the correction factor. 
The smallness of $\theta_{13}$ makes the correction due to $\eem$ and $\eet$ large.}. 

It is evident from Eq.~\ref{eq:rho_l_nsi_maximal} that since the role of 
$\eem$ and $\eet$ are on the same footing, the presence of $\eem$ 
induces an effect identical to that of $\eet$ with the same magnitude. 
But the oscillograms (Figs.~\ref{fig:osc_res_max} and \ref{fig:osc_res_max_neg}) 
for $\eem$ and $\eet$ look quite different. This is because of the fact that 
$\txm$ saturates to a value higher or lower than $90^{\circ}$ in the presence 
of $\eem$ or $\eet$ (see Fig.~\ref{fig:th12_1}). Since $\txm$ is not exactly 
$90^{\circ}$, we have non-zero contributions from some other terms in
$\nu_{\mu} \to \nu_e$ oscillation probability expression, which affect the 
oscillograms in the presence of $\eem$ and $\eet$ in a different fashion.
In the presence of negative $\eem$ and $\eet$, we observe from 
Fig.~\ref{fig:osc_res_max_neg} that we no longer achieve the maximum
transition in $\nu_{\mu} \to \nu_e$ oscillation channel. It is because of the 
fact that in this case, the baseline length required for the maximum 
$\nu_{\mu} \rightarrow \nu_e$ appearance probability turns out 
to be longer than the Earth's diameter (see Eq.~\ref{eq:rho_l_nsi_maximal}).
Therefore, it is not possible to attain the maximum $\nu_{\mu} \rightarrow \nu_e$
transition inside the Earth for negative values of $\eem$ and $\eet$ 
as evident from Fig.~\ref{fig:osc_res_max_neg}. We observe from
Fig.~\ref{fig:osc_res_max} and Fig.~\ref{fig:osc_res_max_neg} 
that in the presence of non-zero $\emt$, there are slight changes 
in $L$ and $E$ as compared to SI case for which we obtain 
maximum possible $\nu_{\mu} \rightarrow \nu_e$ transition.

\section{Impact of NSI  in $\nu_{\mu}\rightarrow\nu_{\mu}$ disappearance channel}
\label{sec:muon-survival-probability}

So far, we have focused on $\nu_{\mu}\rightarrow\nu_e$ appearance 
channel which is one of the most important channel probed in 
LBL experiments. However, another crucial channel, 
$\nu_{\mu}\rightarrow\nu_\mu$ disappearance channel can be probed 
in LBL and atmospheric neutrino experiments. This channel can play 
an important role in precision measurement of the atmospheric oscillation parameters.
In this section, we discuss the effect of NSI in $\nu_{\mu}\rightarrow\nu_{\mu}$ 
survival probability. Since NSI parameters from the (2,3) block have significant 
impact on this channel \cite{Kikuchi:2008vq,Kopp:2007ai}, only these 
NSI parameters have been considered. To get the broad feature, we simplify 
the analysis by assuming $\Delta_{21}\simeq 0$ and $\theta_{13}\simeq 0$.
Under these approximations,  $\nu_{\mu}\rightarrow\nu_{\mu}$ disappearance 
probability expression reduces to~\cite{Mocioiu:2014gua,GonzalezGarcia:2004wg} 
\begin{align}
P_{{\nu_{\mu}\rightarrow\nu_{\mu}}}=1-\sin^22\theta_{23}\sin^2\bigg[\frac{\Delta m^2_{31}L}{4E}\bigg].
\label{eq:surv_1}
\end{align}
Now, we replace the vacuum oscillation parameters by the corresponding modified 
parameters in the presence of SI and NC-NSI assuming the line-averaged constant
Earth matter density. Thus, Eq.~\ref{eq:surv_1} takes the form:
\begin{align}
P^m_{\nu_{\mu}\rightarrow\nu_{\mu}}= 1-\sin^22\theta^m_{23}\sin^2\bigg[\frac{\Delta m^2_{31,m}L}{4E}\bigg].
\label{eq:surv_mat}
\end{align}
Using OMSD approximation ($\Delta m^2_{31}L/4E>>\Delta m^2_{21}L/4E$)  and $\theta_{13}\simeq 0$ in Eq.\ \ref{eq:th23m}, also implementing $\theta_{23}=45^o$, we get
\begin{align}
\sin^22\theta^m_{23}=\frac{(1+2\varepsilon_{\mu\tau}\hat{A})^2}{[(\gamma-\beta)\hat{A}]^2+[1+2\varepsilon_{\mu\tau}\hat{A}]^2}\simeq\bigg[1-\frac{(\gamma-\beta)^2\hat{A}^2}{(1+2\emt\hat{A})^2}\bigg]. 
\label{eq:th23m_omsd}
\end{align}
To calculate $\Delta m^2_{31,m}( = m^2_{3,m}-m^2_{1,m})$ in the last term of Eq.\ \ref{eq:surv_mat}, we use  Eq.\ \ref{eq:m3} and Eq.\ \ref{eq:m1} and implement all the approximations. After simplification, we obtain,
\begin{align}
\Delta m^2_{31,m} = \ldm [\lambda_3-\lambda_2]
\simeq \ldm\bigg[1+2\emt\hat{A}+\frac{1}{2}\frac{(\eff)^2\hat{A}^2}{(1+\emt\hat{A})}\bigg],
\label{eq:m31m}
\end{align}
where, we use the approximation $\theta^m_{12}\rightarrow\pi/2$ in the expression of $m^2_{1,m}$ in Eq.\ \ref{eq:m1}.
So, using Eqs.\ \ref{eq:th23m_omsd} and \ref{eq:m31m}, $\nu_{\mu}\rightarrow\nu_{\mu}$ disappearance probability in presence of NSI parameters from (2,3) sector can be written as,
\begin{align}
P^m_{{\nu_{\mu}\rightarrow\nu_{\mu}}}  = &1-\left[1-\frac{(\gamma-\beta)^2\hat{A}^2}{(1+2\varepsilon_{\mu\tau}\hat{A})^2}\right]\times\sin^2\left[\left\{1+2\varepsilon_{\mu\tau}\hat{A}+\frac{1}{2}\frac{(\gamma-\beta)^2\hat{A}^2}{(1+2\varepsilon_{\mu\tau}\hat{A})}\right\}\frac{\Delta m^2_{31}L}{4E}\right] \nonumber \\
=&\cos^2\left[\left\{1+2\varepsilon_{\mu\tau}\hat{A}+\frac{1}{2}\frac{(\gamma-\beta)^2\hat{A}^2}{(1+2\varepsilon_{\mu\tau}\hat{A})}\right\}\frac{\Delta m^2_{31}L}{4E}\right]\nonumber\\
&\hspace{5cm}+\frac{(\gamma-\beta)^2\hat{A}^2}{(1+2\varepsilon_{\mu\tau})^2}\times\sin^2\left[(1+2\varepsilon_{\mu\tau}\hat{A})\frac{\Delta m^2_{31}L}{4E}\right].
\label{eq:surv_3}
\end{align}
If we only consider the off-diagonal NSI parameter $\emt$, 
the expression boils down to the simplied expression already 
derived in~\cite{Mocioiu:2014gua}. From the approximate expression 
in Eq.~\ref{eq:surv_3}, some broad features about the impact of NSI 
on the $\nu_{\mu}\rightarrow\nu_{\mu}$ survival channel can be observed.
We see that the parameter $(\eff)$ always appears in second order in 
Eq.~\ref{eq:surv_3}, while other NSI parameter $\emt$ has a linear 
dependence. For the same reason, the sign of $(\eff)$, unlike the sign 
of $\emt$, does not affect the disappearance probability. 
Since the strength of NSI parameters are not very large, 
it is expected that the impact of $(\eff)$ will be always 
small compared to $\emt$.

\begin{figure}[tbp]
\vspace{-1cm}
\centering
\includegraphics[scale =0.9]{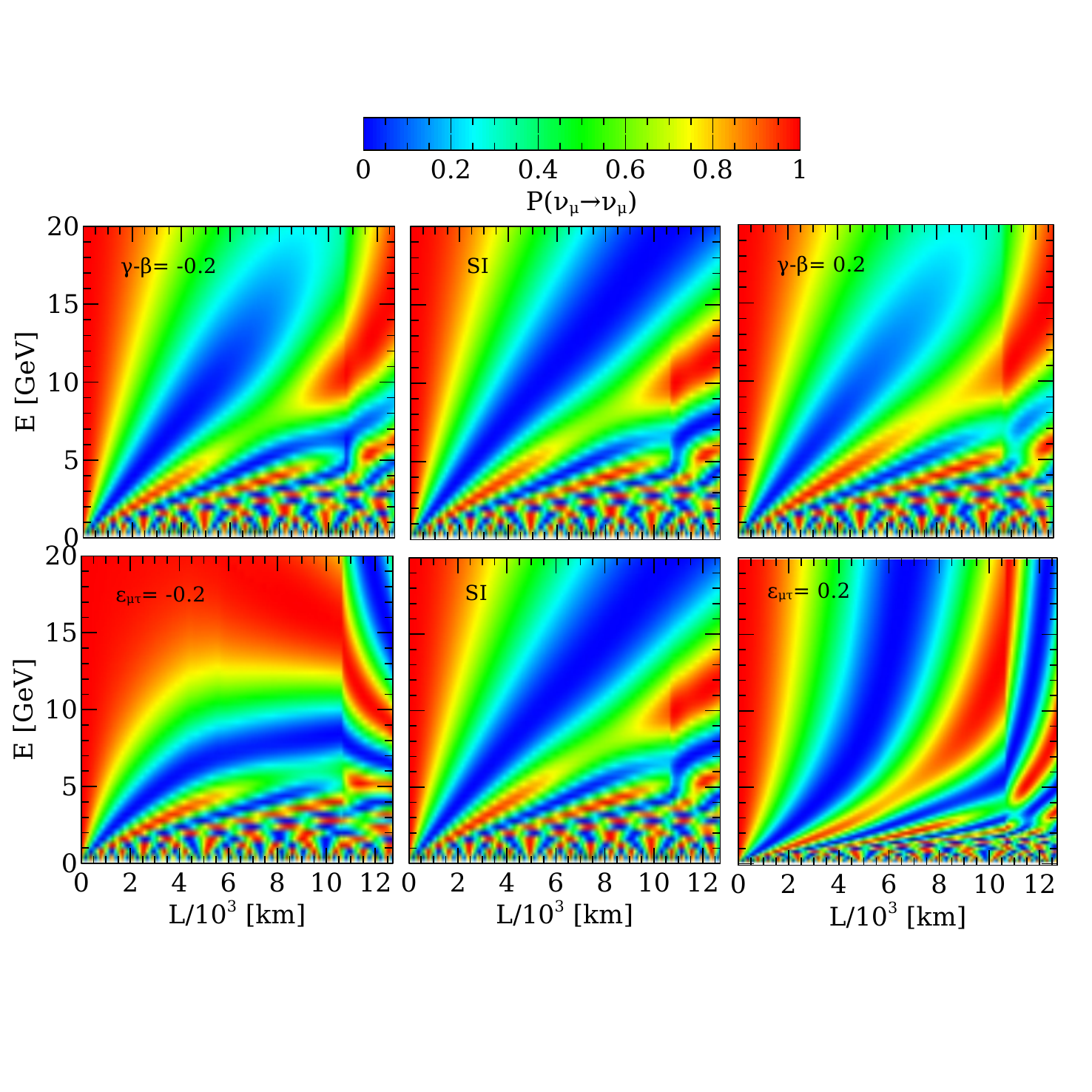}
\vspace*{-1.5cm}
\mycaption{Oscillogram of $\nu_{\mu}\rightarrow\nu_{\mu}$ 
survival probability as a function of baseline $L$ and energy $E$. 
Top and bottom rows correspond to the oscillogram in presence 
of NSI parameter $(\eff)$ and $\varepsilon_{\mu\tau}$, respectively. 
The middle column in both the rows shows the SI case. 	
The first (third) column depicts the presence of negative (positive) 
NSI with a strength of 0.2. Values of the oscillation parameters 
are taken from Table~\ref{table:vac}. We assume 
$\theta_{23}=45^{\circ}$ and NMO.}
\label{fig:surv_1}
\end{figure}

To find out whether these features remain intact even 
if we assume non-zero $\theta_{13}$ and finite $\Delta m^2_{21}$, 
in Fig.\ \ref{fig:surv_1}, we have plotted the $\nu_{\mu}\rightarrow\nu_{\mu}$ 
survival probability as a function of baseline (x-axis) and energy (y-axis) 
commonly known as oscillogram plot. We first consider the full three-flavor 
vacuum expression of  $\nu_{\mu}\rightarrow\nu_{\mu}$ survival probability 
without any approximation~\cite{Agarwalla:2013tza} and replace the vacuum 
parameters with their modified expressions in matter with NSI which we have 
derived in this work. As before, $\theta_{23}$ in vacuum is assumed to be 
$45^{\circ}$. In the top row, we compare the survival probabilities in presence 
of the parameter $(\eff)$ with negative (left column) and positive (right column) 
values and compare with the SI case (middle column). 
We see that there is not any notable variation in oscillation pattern except 
for the magnitude of the survival probability, which decreases at high energies
($E$) and baselines ($L$). Some small differences in the pattern appear 
at some ($L, E$) region. It appears due to non-zero $\theta_{13}$ which 
brings the matter effect into the picture and the finite value of $\Delta_{21}$ 
which gives correction due to solar term. 
As predicted from Eq.~\ref{eq:surv_3}, the sign of $(\eff)$ 
does not affect the $\nu_{\mu} \rightarrow\nu_{\mu}$ disappearance probability. 

In the bottom row of Fig.~\ref{fig:surv_1}, we plot the same but in presence of 
$\emt$ with negative (positive) value in the extreme left (right) panel and show 
the results for SI case in the middle column. It is observed that the presence of 
$\emt$ can lead to significant differences in the pattern of 
$\nu_{\mu}\rightarrow\nu_{\mu}$ disappearance probability as compared 
to SI case. When $\emt$ is positive (negative), one can observe a significant 
shift in the oscillation dip (blue regions emerging from the origin) 
from the SI case towards higher (lower) energies in Fig.~\ref{fig:surv_1}. 
This feature can be explained from the approximate expression 
in Eq.~\ref{eq:surv_3}. In that expression, the value of 
$P^m_{\nu_{\mu} \rightarrow \nu_{\mu}}$ is mainly determined 
by the first term in R.H.S. since second term is suppressed by 
the NSI parameters appearing quadratically. At the first term in 
R.H.S. of Eq.~\ref{eq:surv_3}, minimum occurs at higher (lower) 
energy compared to SI case for a given baseline $L$ when $\emt$ 
is present with positive (negative) strength\footnote{At oscillation dip, 
the argument of the cosine term in Eq.~\ref{eq:surv_3} should be 
approximately equal to $(2n+1)\pi/2$ where $n$ = 0,1,2.... This roughly 
implies that $(1+2\emt\hat{A})\frac{\Delta m^2_{31}L}{4E}\simeq\pi/2$.}. 
Depending on the sign of $\emt$, the regions representing the oscillation 
dip tend to bend upward or downward with increase in baseline length 
compared to SI case.
		 
\section{Summary and Concluding Remarks}
\label{sec:summary}

In this work, we derive the expressions for the evolution 
of the fundamental mass-mixing parameters in the presence 
of SI and SI+NSI considering all possible lepton-flavor-conserving 
and lepton-favor-violating NC-NSI. In order to derive these
expressions, we use a method of approximate diagonalization 
of the effective Hamiltonian by performing successive rotations 
in (2,3), (1,3), and (1,2) blocks. In our study, we present the 
results for the benchmark value of the DUNE baseline of
1300 km and also discuss the results for few other baselines.
We consider both positive and negative values of real NSI
parameters with benchmark values of $\pm~0.2$. 

In the presence of SI only, the 2-3 mixing angle in matter 
($\tzm$) receives a tiny correction which is independent 
of energy and the strength of the matter potential. 
It is observed that only the NSI parameters in the (2,3) block, 
namely $\emt$ and $(\gamma - \beta) \equiv (\ett - \emm)$
influence the evolution of $\tzm$. In the presence of 
negative (positive) value of $(\gamma - \beta$),
$\tzm$ increases (decreases) with energy.
For the maximal value of $\theta_{23}$ in vacuum, 
the change in $\tzm$ is negligible in the presence 
of $\emt$. If $\theta_{23}$ belongs to the upper octant
then $\tzm$ increases (decreases) for negative (positive) 
choices of $\emt$. We notice a completely opposite behavior
if $\theta_{23}$ lies in the lower octant. We also study the 
modification in $\tzm$ as a function of energy when both
the NSI parameters $\emt$ and $(\gamma - \beta)$ 
are present in the scenario with their all possible 
sign combinations. We unravel interesting 
degeneracies in $[\theta_{23}$ - $(\gamma-\beta)]$
and $[\theta_{23}$ - $\emt]$ planes for three
different combination of $L$ and $E$ and discuss 
how our simple approximate analytical expression 
showing the evolution of $\tzm$ plays an important
role to understand these complicated degeneracy
patterns.

In contrast to $\tzm$, $\tym$ is more sensitive in 
matter in the presence of SI and SI+NSI. 
Therefore, an accurate understanding of the
running of $\theta_{13}$ in matter is crucial 
to correctly assess the outcome of the oscillation 
experiments in the presence of NC-NSI. 
$\tym$ goes through an appreciable change 
even in SI case depending on the choice of 
mass ordering and whether we are dealing 
with neutrinos or antineutrinos. 
Compared to SI case, the relative change 
in $\tym$ for ($\nu$, NMO) is somewhat 
suppressed (enhanced) in the presence 
of positive (negative) NSI parameters 
in the (2,3) block, namely 
$\gamma \equiv (\ett - \eee)$, 
$\beta \equiv (\emm - \eee)$, and $\emt$.
For positive $\eem$ and/or $\eet$,
$\tym$ for ($\nu$, NMO) approaches 
the resonance ($\tym = 45^{\circ}$) 
faster than SI case, but after crossing 
the resonance energy, SI takes over. 
For negative $\eem$ or $\eet$,
running of $\tym$ is suppressed almost up to
the resonance energy and then, it increases
very steeply compared to SI case.

As far as the solar mixing angle is concerned, $\txm$ 
approaches to $90^{\circ}$ ($\sin\txm \to 1$, $\cos\txm \to 0$)
very quickly as we increase the neutrino energy in 
SI case. While the NSI parameters in the (2,3) block 
($\gamma, \beta, \emt$) have minimal impact on the running 
of $\txm$, $\eem$ and $\eet$ affect the evolution of $\txm$ 
substantially. In the presence of positive (negative) $\eem$,
the change in $\txm$ with energy qualitatively remains
the same with the saturation value turns out to be 
around $80^{\circ}$ ($100^{\circ}$). In the presence of
positive (negative) $\eet$, the saturation happens 
around $100^{\circ}$ ($80^{\circ}$).

Out of the two mass-squared differences, the evolution of the 
solar $\sdmm$ is quite dramatic as compared to that of the 
atmospheric $\ldmm$ in matter. Both in SI and SI+NSI cases, 
as we increase the energy and go beyond 10 GeV, the value 
of $\sdmm$ increases to almost 20 times as compared to 
its vacuum value for 1300 km baseline. At the same time, the value 
of $\ldmm$ does not change much compared to its vacuum value. 

We demonstrate the utility of our approach in addressing some 
interesting features that we observe in neutrino oscillation in 
presence of matter. It is well known that $\tym$ can attain 
the value of $45^{\circ}$ (MSW-resonance condition) 
for some choices of $L$ and $E$ in the presence of
standard matter effect. Now, in this work, for the first time, 
we show how the $\theta_{13}$-resonance energy gets modified 
in the presence of NC-NSI with the help of simple, compact, 
approximate analytical expressions. We observe that only 
the NSI parameters in the $(2,3)$ block affects the 
$\theta_{13}$-resonance energy. 

We study in detail how the NC-NSI parameters affect 
$\nu_{\mu} \rightarrow \nu_{e}$ oscillation probability
which plays an important role to address the remaining
unknown issues, namely CP violation, mass ordering, and 
the precision measurement of oscillation parameters. 
In this paper, for the first time, we derive a simple 
approximate analytical expression for $E$, $L$ 
and its corresponding $\rho_{\text{avg}}$ to have 
the maximal matter effect which in turn gives rise 
to maximum $\nu_{\mu} \rightarrow \nu_{e}$ transition 
in the presence of all possible NC-NSI parameters.
This analytical expression reveals that in SI case,
maximum $\nu_{\mu} \rightarrow \nu_{e}$ transition 
occurs around the baselines almost passing through 
the core of the Earth ($\simeq$ 10600 km) under the
assumption that the $\theta_{13}$-resonance energy
coincides with the energy that corresponds to the 
first oscillation maximum. However, in the presence 
of positive $\eem$ or $\eet$ in matter, it is observed 
that the baseline length for maximum 
$\nu_{\mu} \rightarrow \nu_{e}$ transition probability 
gets reduced to a much lower value ($L \approx$ 6700 km) 
for the benchmark choices of $\eem$ or $\varepsilon_{e\tau}$ 
equal to 0.2. On the other hand, if we consider negative 
$\eem$ or $\eet$, the required baseline for maximum 
transition probability increases to a value that exceeds
the diameter of the Earth. 

We also study in detail how the NSI parameters 
in the (2,3) block affect $\nu_{\mu} \to \nu_{\mu}$ 
disappearance channel which plays an important role 
in atmospheric neutrino experiments. We observe that the 
off-diagonal NSI parameter $\emt$ has the dominant
effect as compared to the diagonal NSI parameter
$(\gamma-\beta)$. It happens because $(\gamma-\beta)$
appears in the second-order in the approximate
$\nu_{\mu} \rightarrow \nu_{\mu}$ oscillation probability 
expression, whereas $\emt$ shows a liner dependence.  
Also, the sign of $\emt$ has a significant impact on
 $\nu_{\mu} \rightarrow \nu_{\mu}$ disappearance
 channel. However, this oscillation channel is not
 sensitive to the sign of the NSI parameter $(\gamma-\beta)$.
 We hope that the analysis performed in this paper
 will take our understanding of the evolution of the 
 oscillation parameters in the presence of all possible 
 NC-NSI a step forward.

\subsubsection*{Acknowledgments}

We would like to thank T. Takeuchi, P. Denton, and E. Esteban
for useful discussions. S.D. is grateful to the organizers of the
Neutrino 2020 Online Conference at Fermilab, Chicago, USA
during 22nd June to 2nd July, 2020 for giving an opportunity
to present a poster based on this work. S.D. would also like
to thank the organizers of the XXIV DAE-BRNS High Energy
Physics Online Symposium at NISER, Bhubaneswar, India
during 14th to 18th December, 2020 for providing him an
opportunity to give a talk based on this study. We thank the
Department of Atomic Energy (DAE), Govt. of India for
financial support. S.K.A. is supported by the INSPIRE
Faculty Research Grant [IFA-PH-12] from the Department
of Science and Technology (DST), Govt. of India.
S.K.A. acknowledges the financial support from the
Swarnajayanti Fellowship Research Grant
(No. DST/SJF/PSA-05/2019-20) provided by
the DST, Govt. of India and the Research Grant
(File no. SB/SJF/2020-21/21) from the Science
and Engineering Research Board (SERB) under
the Swarnajayanti Fellowship by the DST, Govt. of India.
S.K.A. and M.M. acknowledge the financial support
from the Indian National Science Academy (INSA)
Young Scientist Project [INSA/SP/YS/2019/269].
M.M. acknowledges the support of IBS under
the project code IBS-R018-D1.

\bibliographystyle{JHEP}
\bibliography{reference}

\end{document}